\documentclass[12pt]{article}
\usepackage{a4wide}
\usepackage{chicago}
\usepackage{color}
\usepackage{graphics}
\usepackage{subfigure}
\usepackage[normalem]{ulem}
\usepackage{epstopdf}
\usepackage{epsfig}
\usepackage{lscape}

\newcounter{appeneqnA}%

\newcounter{appeneqnB}%

\newcounter{appeneqnC}%

\newcounter{appeneqnD}%

\begin{document}

\noindent {\bf Title: Airfall on Comet 67P/Churyumov--Gerasimenko}

\bigskip\bigskip

\begin{trivlist}
\item {\bf Author (1):} Bj\"{o}rn J. R. Davidsson
\item {\bf Affiliation:} Jet Propulsion Laboratory, California Institute of Technology
\item {\bf Address:} M/S 183--401, 4800 Oak Grove Drive, Pasadena, CA 91109, USA
\item {\bf Telephone:} (818)--393--8032
\item {\bf E--mail address:} bjorn.davidsson@jpl.nasa.gov
\end{trivlist}

\bigskip

\begin{trivlist}
\item {\bf Author (2):} Samuel Birch
\item {\bf Affiliation:} Cornell University
\item {\bf Address:} 426 Space Sciences Building, Cornell University, Ithaca NY, 14850, USA
\item {\bf Telephone:} (510)--712--0270
\item {\bf E--mail address:} sb2222@cornell.edu
\end{trivlist}

\bigskip

\begin{trivlist}
\item {\bf Author (3):} Geoffrey A. Blake
\item {\bf Affiliation:}  Division of Geological \& Planetary Sciences, California Institute of Technology
\item {\bf Address:} MC150--21, Pasadena, CA 91125, USA
\item {\bf Telephone:}  (626)--395--6296
\item {\bf E--mail address:} gab@gps.caltech.edu
\end{trivlist}

\bigskip

\begin{trivlist}
\item {\bf Author (4):} Dennis Bodewits
\item {\bf Affiliation:} Department of Physics, Auburn University
\item {\bf Address:} 201 Allison Laboratory, Auburn, AL 36849, USA
\item {\bf Telephone:} (334)--844--4274
\item {\bf E--mail address:} dennis@auburn.edu
\end{trivlist}

\bigskip

\begin{trivlist}
\item {\bf Author (5):} Jason P. Dworkin
\item {\bf Affiliation:} NASA Goddard Space Flight Center
\item {\bf Address:} Greenbelt, MD 20771, USA
\item {\bf Telephone:} (301)--286--8631
\item {\bf E--mail address:} jason.p.dworkin@nasa.gov
\end{trivlist}

\bigskip
\begin{trivlist}
\item {\bf Author (6):} Daniel P. Glavin
\item {\bf Affiliation:} NASA Goddard Space Flight Center
\item {\bf Address:} Greenbelt, MD 20771, USA
\item {\bf Telephone:} (301)--614--6361
\item {\bf E--mail address:} daniel.p.glavin@nasa.gov
\end{trivlist}

\bigskip

\begin{trivlist}
\item {\bf Author (7):}  Yoshihiro Furukawa
\item {\bf Affiliation:} Department of Earth Science, Tohoku University
\item {\bf Address:} 6--3, Aza--aoba, Aramaki, Aoba--ku, Sendai 980--8578, Japan
\item {\bf Telephone:} +81-22-795-3453
\item {\bf E--mail address:} furukawa@tohoku.ac.jp
\end{trivlist}

\bigskip

\begin{trivlist}
\item {\bf Author (8):} Jonathan I. Lunine
\item {\bf Affiliation:} Department of Astronomy and Carl Sagan Institute, Cornell University
\item {\bf Address:} Cornell University, Ithaca, NY 14850, USA
\item {\bf Telephone:} (607)--255--5911
\item {\bf E--mail address:} jlunine@astro.cornell.edu
\end{trivlist}

\bigskip

\begin{trivlist}
\item {\bf Author (9):} Julie L. Mitchell
\item {\bf Affiliation:} NASA Johnson Space Center
\item {\bf Address:} 2101 NASA Pkwy, Houston, TX 77058, USA
\item {\bf Telephone:} (281)--483--2925
\item {\bf E--mail address:} julie.l.mitchell@nasa.gov
\end{trivlist}

\bigskip

\begin{trivlist}
\item {\bf Author (10):} Ann N. Nguyen
\item {\bf Affiliation:} Jacobs, Houston TX; Astromaterials Research and Exploration Science, NASA Johnson Space Center, Houston TX
\item {\bf Address:} 2101 NASA Parkway, Mail Code XI3, Houston TX 77058
\item {\bf Telephone:} 281--483--9446
\item {\bf E--mail address:} lan-anh.n.nguyen@nasa.gov
\end{trivlist}

\bigskip

\begin{trivlist}
\item {\bf Author (11):} Steve Squyres
\item {\bf Affiliation:} Blue Origin, LLC
\item {\bf Address:} 21218 76th Ave S, Kent, WA 98032, USA
\end{trivlist}

\bigskip

\begin{trivlist}
\item {\bf Author (12):} Aki Takigawa
\item {\bf Affiliation:} The Hakubi Center for Advanced Research / Division of Earth and Planetary Science
\item {\bf Address:} Kyoto University, Kitashirakawa--Oiwakecho, Sakyo, Kyoto 606--8502, Japan
\item {\bf Telephone:} +81--75--753--4169
\item {\bf E--mail address:} takigawa@kueps.kyoto-u.ac.jp
\end{trivlist}

\bigskip

\begin{trivlist}
\item {\bf Author (13):} Jean--Baptiste Vincent
\item {\bf Affiliation:} DLR Insitute of Planetary Research
\item {\bf Address:} Rutherfordstrasse 2, 12489 Berlin, Germany
\item {\bf Telephone:} +49--30--67055--7912
\item {\bf E--mail address:} jean-baptiste.vincent@dlr.de
\end{trivlist}

\bigskip

\begin{trivlist}
\item {\bf Author (14):} Kris Zacny
\item {\bf Affiliation:} Honeybee Robotics
\item {\bf Address:} 398 W Washington Blvd., Pasadena, CA 91103
\item {\bf Telephone:} 646--508--9807
\item {\bf E--mail address:} kazacny@honeybeerobotics.com
\end{trivlist}

\bigskip

\begin{trivlist}
\item {\bf Number of manuscript pages: }65
\item {\bf Number of figures: }11
\item {\bf Number of tables: }4
\end{trivlist}

\newpage

\noindent {\bf Running head: }Airfall on Comet 67P/Churyumov--Gerasimenko

\begin{trivlist}
\item {\bf Address for editorial correspondence:}
\item Bj\"{o}rn Davidsson
\item Jet Propulsion Laboratory
\item  M/S 183--401  
\item 4800 Oak Grove Drive
\item Pasadena, CA 91109 
\item USA
\end{trivlist}

\noindent
Email: bjorn.davidsson@jpl.nasa.gov\\

\noindent
(818)--393--8032 (phone)

\newpage

\begin{abstract}

\noindent
We here study the transfer process of material from one hemisphere to the other (deposition of airfall material) on an active comet nucleus, 
specifically 67P/Churyumov--Gerasimenko. Our goals are to: 1) quantify the thickness of the airfall debris layers and how it depends on the location of 
the target area, 2) determine the amount of $\mathrm{H_2O}$ and $\mathrm{CO_2}$ ice that are lost from icy dust assemblages of different sizes during 
transfer through the coma, and 3) estimate the relative amount of vapor loss in airfall material after deposition in order to understand what locations 
are expected to be more active than others on the following perihelion approach.\\

\noindent
We use various numerical simulations, that include orbit dynamics, thermophysics of the nucleus and of individual coma aggregates, coma 
gas kinetics and hydrodynamics, as well as dust dynamics due to gas drag, to address these questions. We find that the thickness of accumulated airfall material varies 
substantially with location, and typically is of the order $0.1$--$1\,\mathrm{m}$. The airfall material preserves substantial amounts of water 
ice even in relatively small (cm--sized) coma aggregates after a rather long ($12\,\mathrm{h}$) residence in the coma. However, $\mathrm{CO_2}$ is 
lost within a couple of hours even in relatively large (dm--sized) aggregates, and is not expected to be an important component in airfall deposits. We 
introduce reachability and survivability indices to measure the relative capacity of different regions to simultaneously collect airfall and to preserve its water ice until the 
next perihelion passage, thereby grading their potential of contributing to comet activity during the next perihelion passage. 

\bigskip

\end{abstract}
\newpage

\section{Introduction} \label{sec_intro}

The initial characterization of Comet 67P/Churyumov--Gerasimenko (hereafter 67P) by the OSIRIS cameras \shortcite{kelleretal07} 
on the Rosetta orbiting spacecraft \shortcite{glassmeieretal07} revealed widespread smooth terrains on the northern hemisphere of the nucleus, 
primarily in Ma'at on the small lobe, in Ash, Anubis, and Imhotep on the large lobe, and in Hapi that constitutes the northern neck region between the two lobes 
(\shortciteNP{sierksetal15}; \shortciteNP{thomasetal15a}). The smooth material in the Ma'at and Ash regions formed a relatively thin coverage over 
partially--revealed consolidated landforms. This led \shortciteN{thomasetal15a} to propose that the smooth material, at least in those regions, 
constituted a rather recent veneer of dust that had rained down from the coma. They named these deposits and the process forming them ``airfall''. 
The presence of smooth deposits in isolated topographic lows, e.g., in Khepry \shortcite{elmaarryetal15}, or in large gravitational lows found primarily 
in the Imhotep and Hapi regions, suggest that airfall may not be the only mechanism responsible for the formation of smooth terrain. \shortciteN{augeretal15a} propose 
that the vast smooth plain in Imhotep formed through mass wasting the surrounding steep cliffs and by transport downhill toward the lowland. 
Mass wasting from cliffs revealed by taluses, gravitational accumulation deposits, and diamictons that extends into Hapi \shortcite{pajolaetal19} indicate 
that such processes are partially responsible for the smooth material in that region as well. The presence of deposit--free regions that sharply contrast with surrounding 
smooth terrain, best illustrated by the Aten depression on the large lobe or the Anuket region in the neck area  \shortcite{elmaarryetal15}, suggests that the removal 
rate of airfall material through self--cleaning is locally high  and that the net accumulation rate may be slow (e.g., in case the Aten depression formed 
recently in a massive outburst event).\\

\noindent
At the time of the Rosetta orbit insertion around 67P in August 2014 the northern hemisphere was illuminated while the southern hemisphere 
experienced polar night because of the orientation of the spin axis \shortcite{sierksetal15}. Most activity, as revealed by prominent dust jets, came 
from Hapi (\shortciteNP{sierksetal15}; \shortciteNP{laraetal15}), that also was the brightest and bluest unit in terms of spectral slope \shortcite{fornasieretal15}, 
showing that the neck region had accumulated particularly ice--rich material that was strongly active. \shortciteN{thomasetal15a} noted that the airfall 
deposits on north--facing regions are more extensive than on south--facing regions and proposed that the major transport route went from Hapi to other 
suitably oriented regions on the northern hemisphere. During approach to the inbound equinox in May 2015, increasingly large portions of the southern 
hemisphere were illuminated and eventually received peak solar illumination at the perihelion passage in August 2015 while the northern hemisphere 
experienced polar night. \shortciteN{kelleretal15} demonstrated that $\sim 80\%$ of the solar radiation is absorbed by the northern hemisphere at a 
low rate over an extended amount of time, while the southern hemisphere receives the remaining $\sim 20\%$ during a much shorter time interval. 
Their thermophysical modeling showed that the strongly non--linear response of sublimation to solar flux levels led to a $\sim 3$ times higher 
erosion in the south compared to the north. They therefore suggested that the main transport route of airfall material went from the southern 
hemisphere to the northern hemisphere. The views of \shortciteN{thomasetal15a} and \shortciteN{kelleretal15} are not in contradiction, but suggest that 
activity in the south around perihelion leads to airfall in the cold and inactive north. Some of this material is then redistributed from 
Hapi to other northern regions from the time of the outbound equinox to the shutdown of activity at larger heliocentric distance, with a pause during the 
aphelion passage followed by resumed redistribution when activity reappears during inbound motion.\\

\noindent
OSIRIS observations of the southern hemisphere showed a strong hemispherical dichotomy regarding smooth terrains -- they were essentially 
missing in the south (\shortciteNP{elmaarryetal16}; \shortciteNP{leeetal16}; \shortciteNP{birchetal17}), showing that any deposited or otherwise 
produced smooth material is rapidly cleaned off. Observations by the mass spectrometer ROSINA (\shortciteNP{hassigetal15}; \shortciteNP{fougereetal16}) and the 
near--infrared spectrometer VIRTIS \shortcite{finketal16} on Rosetta show that the two hemispheres display a strong chemical dichotomy as well. The northern hemisphere is predominantly outgassing $\mathrm{H_2O}$ and 
comparably small amounts of $\mathrm{CO_2}$, while the southern hemisphere is a source of both water and carbon dioxide. \shortciteN{kelleretal17} interpreted the 
chemical dichotomy as a result of airfall. In their view, the $\mathrm{CO_2}$ is either missing in the solid material being ejected into the coma from the south near perihelion 
(i.e., the $\mathrm{CO_2}$ sublimation front is located at some depth) or is lost on the way during transport toward the north. Because of the substantially 
higher sublimation temperature of $\mathrm{H_2O}$ compared to $\mathrm{CO_2}$ the water loss is significantly smaller. The airfall that is responsible for 
northern activity in other parts of the orbit is therefore rich in water ice but poor in supervolatiles like $\mathrm{CO_2}$, according to \shortciteN{kelleretal17}.\\

\noindent
High--resolution imaging of smooth terrain in Ash by OSIRIS during low ($\sim 10\,\mathrm{km}$) Rosetta orbits \shortcite{thomasetal15b}, by the Philae/ROLIS 
camera during descent toward Agilkia in Ma'at (\shortciteNP{mottolaetal15}; \shortciteNP{pajolaetal16}), and by OSIRIS during the landing of Rosetta at 
Sais in Ma'at \shortcite{pajolaetal17b} revealed that the material in those locations consisted primarily of pebbles, cobbles, and boulders in the cm--m 
size range. In the following, such units are collectively referred to as ``chunks''.  The observations of similarly sized chunks in the coma (e.g., \shortciteNP{rotundietal15}; 
\shortciteNP{davidssonetal15b}), of which a substantial fraction display acceleration toward the nucleus \shortcite{agarwaletal16}, show that the airfall concept is 
viable.\\

\noindent
Smooth terrains are not featureless. Structures resembling aeolean ripples were observed in Hapi, dunes that in some cases contained pits (potentially formed by sublimation) 
were seen in Serqet and Maftet, and some boulders in Hapi and Maftet appeared to have wind tails \shortcite{thomasetal15b}, that are also seen in Ma'at \shortcite{mottolaetal15}. 
Whether such features primarily form during airfall deposition or are the result of lateral transport mechanisms is unclear. They do indicate that deposition is not a trivial phenomenon 
and that significant local transport may take place after deposition. The features seen in smooth terrain are also not static. The first reported observations 
of large--scale morphological changes concerned roundish scarps that formed and expanded at $\sim 6\,\mathrm{m\,day^{-1}}$ in Imhotep \shortcite{groussinetal15b}. 
Later, similar scarp retreats were also seen in Hapi, Anubis, and in Seth (\shortciteNP{elmaarryetal17}; \shortciteNP{huetal17}). The scarps frequently displayed strong brightening 
and spectral slope changes suggestive of the presence of abundant sub--surface water ice that mixed up to the surface during the retreat process. A dramatic and puzzling 
example of these morphological changes where the removal of the aeolean ripples in Hapi in April--July 2015 by retreating scarps, that was followed by ripple reformation in 
December 2015 \shortcite{elmaarryetal17}.\\

\noindent
OSIRIS observations allowed for a documentation of inbound removal and outbound redeposition of airfall material, albeit restricted to specific locations and 
time instances because of resolution, viewing geometry, and imaging cadence restrictions. A comprehensive summary is found in \shortciteN{huetal17}. 
The presence of ``honeycomb'' features in Ash, Babi, Serqet, Seth, and particularly in Ma'at in late March 2015 (also see \shortciteNP{shietal16}) caused 
by the removal of overlying granular material is indicative of self--cleaning. The time--scale of their formation is uncertain because of observational 
biases but the earliest known honeycomb sighting occurred two months earlier. Other signs of airfall deposit removal include exposure of previously 
hidden outcrops and boulders, formation of shallow depressions, and smoothing of previously pitted deposits \shortcite{huetal17}. The OSIRIS images 
were used to create three--dimensional maplets of selected terrains before and after the removal of deposits. Although it is not possible to estimate the 
thickness of the expelled layers by simply subtracting two maplets, insight about the erosion can still be obtained by measuring changes in the 
degree of roughness inferred from such maplets. \shortciteN{huetal17} find, for regions where changes are more easily detected, that at least $0.5\,\mathrm{m}$ 
has been removed (similar to the pixel resolution) and an estimated average removal of $1.3\,\mathrm{m}$ for one specific honeycomb. Post--perihelion observations showed that the 
honeycombs had disappeared and were replaced by smooth terrain. \shortciteN{huetal17} do not quantify the thickness of the redeposited layer but a reasonable assumption 
is of order $0.1$--$1\,\mathrm{m}$, so that the net orbital change is small. It is not self--evident that the northern airfall deposits currently 
experience net growth, or if there is currently net erosion of deposits built up at an earlier time when conditions were different than today.\\

\begin{figure}
\begin{center}
     \scalebox{0.6}{\includegraphics{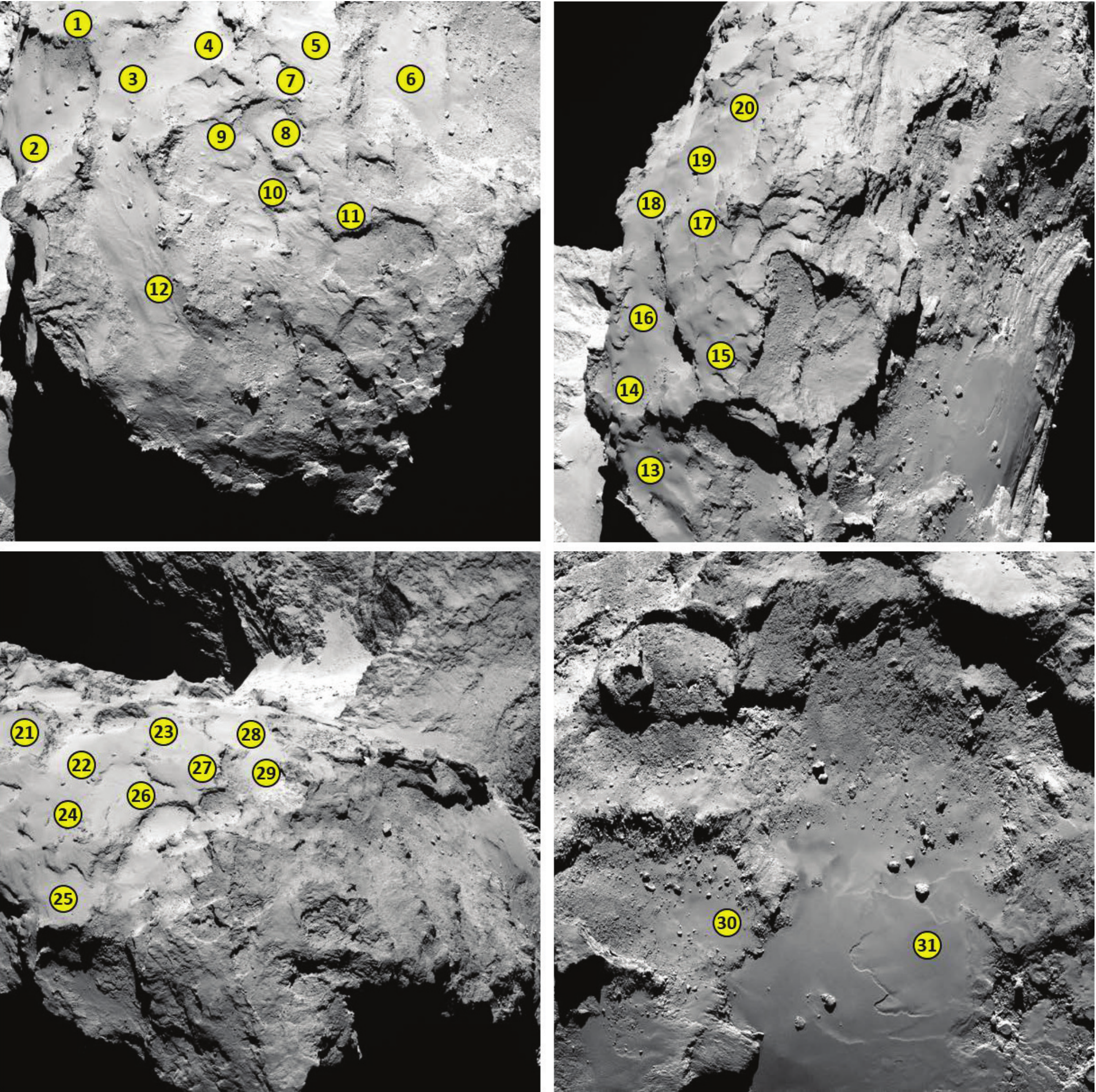}}
   \caption{The location of target areas, considered for CAESAR sample collection, on Rosetta/OSIRIS context images with file names provided in the following. Upper left: Ma'at (N20140902T084254618ID30F22.img). Upper right: eastern Ash 
(N20140822T054254583ID30F22.img). Lower left: western Ash (N20140902T121022552ID30F22.img). Lower right: Imhotep (N20140905T064555557ID30F22.img).}
     \label{fig0}
\end{center}
\end{figure}

\noindent
From the description above a basic qualitative understanding has emerged regarding the origin and behavior of smooth 
terrain on 67P; this new insight emphasizes the global redistribution of material on the nucleus enabled by its spin axis obliquity. This process creates a 
non--primordial chemical surface heterogeneity. If these processes are active on other comet nuclei as well it  makes the interpretation of 
groundbased production rate patterns more difficult than previously envisioned. However, it is also evident that a detailed understanding of these processes requires substantial modeling efforts that can fill the gaps that 
are not immediately provided by available observations.  Therefore, the first goal of this paper is to present a quantitative study of the south--to--north 
transfer process proposed by \shortciteN{kelleretal15} and elaborated on by \shortciteN{kelleretal17}, i.e., to estimate the thickness of the deposited 
layer built during the perihelion passage and how it varies across the northern hemisphere. Specifically, we focus on 31 target areas distributed within the 
Ma'at region on the small lobe and within the Ash and Imhotep regions on the large lobe. These sites were selected and investigated in the context of a Phase~A study 
for the New~Frontiers~4 candidate mission CAESAR (Comet Astrobiology Exploration SAmple Return). The sites were selected to cover a range of regions of smooth terrain in 
the northern hemisphere of 67P that were deemed to be accessible to the CAESAR touch and go sampling. The location of the 31 target areas are shown in Fig.~\ref{fig0}.\\

\noindent
Our second goal is to quantify the loss of volatiles expected to take place during the transfer process, thereby enabling a comparison between the ice abundance of 
fresh airfall deposits with that of the sources of this material. Our third goal is to quantify the relative capacity of different airfall deposition sites 
to retain their water ice content after deposition, in order to better understand their capability to uphold comet activity as the comet approaches the 
Sun after the aphelion passage. In the context of the CAESAR mission this work contributed to locating high science value sampling 
sites in the smooth terrain on 67P for future Earth return.\\

\noindent
Our work relies on a pipeline of different models and methods that are described in Sec.~\ref{sec_method}. In brief, we first find Keplerian orbits 
that connect specific source regions and target areas that are unobstructed by the rotating nucleus. Those orbits are characterized by a specific 
ejection velocity $v_{\rm d}$ that needs to be provided by gas drag, and are realized for a specific dust diameter $a_{\rm crit}$. We calculate $a_{\rm crit}$ 
by first determining the local surface temperature  and outgassing rate, use that information to calculate the coma number density, translational temperature, 
and gas expansion velocity versus height, which in turn are used to evaluate the gas drag force needed for dust dynamics calculations. We also present our 
model for calculating the loss of $\mathrm{H_2O}$ and $\mathrm{CO_2}$ from chunks in the coma, as well as the longterm nucleus thermophysics modeling. 
Our results are presented in Sec.~\ref{sec_results}, we discuss these results and compare with previously published airfall investigations in Sec.~\ref{sec_discussion}, 
and summarize our conclusions in Sec.~\ref{sec_conclusions}.

\section{Methods} \label{sec_method}

Section~\ref{sec_method_kepler} describes the method for determining whether an unobstructed Keplerian orbit exists that 
connects a certain southern source region with a specific northern target area, taking the irregular shape of the rotating nucleus 
into account. The purpose of that calculation is to identify the dust ejection velocity $v_{\rm d}$ that is needed in order to realize a particular 
orbit. The reliability of this analytical method based on the point--mass assumption is tested using numerical integration of 
orbits around irregular bodies with complex gravity fields described in Sec.~\ref{sec_method_numorb}.\\ 

\noindent
The thermophysical model described in Sec.~\ref{sec_method_nucthermo} is needed in order to calculate the nucleus outgassing 
rate as function of surface location and time. The outgassing rate is used to calculate how the gas density, 
drift (expansion) speed, and temperature vary with height, using procedures described in Sec.~\ref{sec_method_coma}. Those quantities 
are, in turn, necessary to determine the specific chunk size that will be accelerated to the particular $v_{\rm d}$ needed to 
reach a given target area, as summarized in Sec.~\ref{sec_method_eject}. With all of this information in hand, the amount of airfall material 
can be calculated, as described in Sec.~\ref{sec_method_airfall}.\\

\noindent
Section~\ref{sec_method_grainthermo} describes the model used for modeling the thermophysical evolution of chunks in the coma, 
and the longterm thermophysical evolution of target sites are described in Sec.~\ref{sec_method_volatileloss}. The mathematical 
symbols used throughout this work are summarized in Tables~\ref{tab0}--\ref{tab0b}.

\begin{table}
\begin{center} {\bf Model parameters \#1} \end{center}
\begin{center}
\footnotesize
\begin{tabular}{||l|l|l||}
\hline
\hline
Symbol & Description & Unit\\
\hline
$A$ & Bond albedo & --\\
$\mathcal{A}$ & Chunk cross section & $\mathrm{m^2}$\\
$\mathcal{A}_{\rm crit}$ & Cross section of maximum liftable chunk & $\mathrm{m^2}$\\
$\mathcal{C}$ & Eq.~(\ref{eq:25}) constant & $\mathrm{K\,(1-\gamma)^{-1}}$\\
$C_{\rm D}$ & Drag coefficient & --\\
$D$ & Dust chunk diameter & $\mathrm{m}$\\
$D_{\rm min},\,D_{\rm max},\,D_{\rm c}$ & Minimum, maximum, and intermediate dust chunk diameter & $\mathrm{m}$\\
$\mathcal{D}$ & Molecular number flux & $\mathrm{m^{-2}\,s^{-1}}$\\
$E_{\rm t}$ & Molecular kinetic energy & $\mathrm{J}$ \\
$E_{\rm r}$ & Molecular rotational energy & $\mathrm{J}$\\
$F_{\rm drag}$ & Drag force & $\mathrm{N}$\\
$\mathbf{F}_{\rm g}$ & Gravitational force vector & $\mathrm{N}$\\
$F_{\imath\jmath}$ & Fraction of latent heat consumed during segregation & --\\
$\mathcal{F}_{jk}$ & View factor & --\\
$\mathcal{F}$ & Fraction of dust mass at $D>D_{\rm c}$ & --\\
$F,\,G$ & Eq.~(\ref{eq:18}) parameters & --\\
$G_{\rm N}$ & Newtonian gravitational constant & $\mathrm{N\,m^2\,kg^{-2}}$\\  
$H_{\imath}$ & Energy release during crystallization & $\mathrm{J\,kg}$\\
$K_{\rm t}$ & Knudsen layer thickness & $\mathrm{m}$\\
$\mathcal{L}$ & Latent heat of sublimation & $\mathrm{J\,kg^{-1}}$\\
$L'$ & Normalized relative ice loss of target area & --\\
$M,\,M_1,\,M_2$ & Mach number & --\\
$M_{\rm n}$ & Comet nucleus mass & $\mathrm{kg}$\\
$M_{\rm rot}$ & Rotation matrix & --\\
$P$ & Nucleus or chunk rotation period & $\mathrm{h}$\\
$\mathcal{P}$ & Comet orbital period & $\mathrm{yr}$\\
$Q_{\rm d}$ & Dust production rate & $\mathrm{kg\,m^{-2}\,s^{-1}}$\\
$Q_j,\, Q_{\rm s}$ & Water sublimation rate & $\mathrm{kg\,m^{-2}\,s^{-1}}$\\
$R_{\rm n}$ & Comet nucleus curvature radius & $\mathrm{m}$\\
$R_*$ & Radiogenic energy production rate & $\mathrm{J\,m^{-3}\,s^{-1}}$\\
$\mathcal{R}$ & Reachability index & --\\
$S_1$ & Speed ratio & --\\
$S_{\odot}$ & Solar constant & $\mathrm{J\,m^{-2},s^{-1}}$\\
$S'$ & Normalized relative ice retainment capacity of target area & --\\
$\mathcal{S}$ & Survivability index & --\\
$T,\,T_j(x,t),\,T_k(x,t),\,T_{\rm s},\,T_0,\,T_1,\,T_{\infty}$ & Temperature & $\mathrm{K}$\\
$T_{\rm fr}$ & Freeze--out temperature & $\mathrm{K}$\\
$T_{\rm r}$ & Rotational temperature & $\mathrm{K}$\\
$T_{\rm surf}$ & Surface temperature & $\mathrm{K}$\\
$U$ & Gravitational potential & $\mathrm{J\,kg^{-1}}$\\
$W,\,W_1,\,W_2,\,W_{\infty}$ & Gas drift speed & $\mathrm{m\,s^{-1}}$\\
$\mathcal{W}$ & Speed ratio & --\\
$Z$ & Hertz--Knudsen formula & $\mathrm{kg\,m^{-2}\,s^{-1}}$\\
\hline 
\hline
\end{tabular}
\caption{Functions and parameters with descriptions and units. Upper case Latin.}
\label{tab0}
\end{center}
\end{table}

\begin{table}
\begin{center} {\bf Model parameters \#2} \end{center}
\begin{center}
\footnotesize
\begin{tabular}{||l|l|l||}
\hline
\hline
Symbol & Description & Unit\\
\hline
$a,\,b,\,c$ & Eq.~(\ref{eq:05}) coefficients  & $\mathrm{m^3\,s^{-1}}$\\
$a_1,\,a_2,\,a_3$ & Eq.~(\ref{eq:11}) coefficients & --\\
$a_{\rm crit}$ & Radius of maximum liftable chunk & $\mathrm{m}$\\
$a_{\rm o}$ & Orbital semi--major axis & $\mathrm{m}$\\
$c_{\rm s}$ & Specific heat capacity & $\mathrm{J\,kg^{-1}\,K^{-1}}$\\
$d$ & Distance between chunk and target location & $\mathrm{m}$\\
$f$ &  Eq.~(\ref{eq:05}) coefficient  & --\\
$f_i$ & Ice volume fraction & --\\
$g_{\imath}$ & Heat capacity of gaseous species $\imath$ & $\mathrm{J\,kg^{-1}}$\\
$h$ & Height above the comet nucleus surface & $\mathrm{m}$\\
$\mathbf{h}=\{h_{\rm x},\,h_{\rm y},\,h_{\rm z}\}$ & Chunk orbit normal vector & $\mathrm{m^2\,s^{-1}}$\\
$\imath$ & Chemical species identifier (host) & --\\
$\jmath$ & Chemical species identifier (guest) & --\\
$j,\,k$ & Facet number & --\\
$k_{\rm B}$ & Boltzmann constant & $\mathrm{J\,K^{-1}}$\\
$l$ & Spherical chunk latitudinal coordinate & $\mathrm{rad}$\\
$m$ & Water molecule mass & $\mathrm{kg}$\\
$m_{\imath}$ & Molecule mass of species $\imath$ & $\mathrm{kg}$\\
$m_{\rm d}$ & Mass of a dust chunk & $\mathrm{kg}$\\
$n,\,n_{\rm s},\,n_0,\,n_1,\,n_2$ & Gas number density & $\mathrm{m^{-3}}$\\
$n_{\rm sp}$ & Number of chemical species & --\\
$\hat{n}_{\rm s}$ & Surface normal unit vector & --\\
$p,\, q$  & Eq.~(\ref{eq:05}) coefficients  & --\\
$p_{\imath}$ & Partial pressure of species $\imath$ & $\mathrm{Pa}$\\
$q_{\imath}$ & Volume mass production rate of species $\imath$ & $\mathrm{kg\,m^{-3}\,s^{-1}}$\\
$q'_{\imath}$ & Volume mass segregation or crystallization rate & $\mathrm{kg\,m^{-3}\,s^{-1}}$\\
$p_{\rm sat}$ & Water saturation pressure & $\mathrm{Pa}$\\
$p_{\rm v}$ & Vapor pressure & $\mathrm{Pa}$\\
$q_{\rm s}$ & Dust chunk size distribution power--law index & --\\
$r$ & Spherical chunk radial coordinate & $\mathrm{m}$\\
$\mathbf{r}_{\rm d}$ & Chunk position vector & $\mathrm{m}$\\
$r_{\rm g}$ & Constituent grain radius & $\mathrm{m}$\\
$r_{\rm h}$ & Heliocentric distance & $\mathrm{AU}$\\
$r_{jk}$ & Distance between facets & $\mathrm{m}$\\
$r_{\rm o}$ & Orbital distance between chunk and nucleus center & $\mathrm{m}$\\
$\mathbf{r}_{\rm s}$ & Source region position vector & $\mathrm{m}$\\
$\mathbf{r}_{\rm t}=\{r_{\rm tx},\,r_{\rm ty},\,r_{\rm tz}\}$ & Target area inertial position vector & $\mathrm{m}$\\
 & with $\mathbf{r}_{\rm t}'=\mathbf{r}_{\rm t}(t=0)$ & \\
$s_k$ & Facet area & $\mathrm{m^2}$\\
$t$ & Time & $\mathrm{s}$\\
$t_{\rm pl}$ & Target area crossing time with chunk orbital plane & $\mathrm{s}$\\
$u_{\rm d}$ & Chunk speed & $\mathrm{m\,s^{-1}}$\\
$v_{\rm d}$ & Post--acceleration speed along surface normal & $\mathrm{m\,s^{-1}}$\\
$v_{\rm esc}$ & Escape velocity & $\mathrm{m\,s^{-1}}$\\
$v_{\rm o}$ & Orbital speed & $\mathrm{m\,s^{-1}}$\\
$v_{j\odot},\,v_{k\odot}$ &  Sunlit fraction of facet & --\\
$\mathbf{v}_{\rm s}$ & Dust chunk ejection velocity vector & $\mathrm{m\,s^{-1}}$\\
$x,\,x_j$ & Depth below the comet surface & $\mathrm{m}$\\
$\{\hat{x},\,\hat{y},\,\hat{z}\}$ & Inertial system coordinate system unit vectors & --\\
 & ($\hat{z}$ parallel to the nucleus rotation axis) & \\
\hline 
\hline
\end{tabular}
\caption{Functions and parameters with descriptions and units. Lower case Latin.}
\label{tab0a}
\end{center}
\end{table}

\begin{table}
\begin{center} {\bf Model parameters \#3} \end{center}
\begin{center}
\footnotesize
\begin{tabular}{||l|l|l||}
\hline
\hline
Symbol & Description & Unit\\
\hline
$\alpha$ & Right ascension & $^{\circ}$\\
$\alpha_{\rm bf}$ & Molecular surface backflux fraction & --\\
$\alpha_{\rm s}$ & Sublimation coefficient & --\\
$\beta_j,\,\beta_k$ & Angle between facet normal and direction to other facet & $\mathrm{rad}$\\
$\beta^-$ & Molecular backflux ratio & --\\
$\gamma$ & Gas heat capacity ratio &  --\\
$\delta$ & Declination & $^{\circ}$\\
$\varepsilon$ & Emissivity & --\\
$\zeta$ & Number of rotational degrees of freedom & --\\
$\kappa$ & Heat conductivity & $\mathrm{W\,m^{-1}\,K^{-1}}$\\
$\lambda,\,\lambda_0$ & Molecular mean free path & $\mathrm{m}$\\
$\mu_j,\,\mu_k$ & Cosine of incidence angle & --\\
$\mu_{\rm o},\,\mu_{\rm}^*$ & Reduced mass & $\mathrm{kg}$\\
$\rho$ & Density & $\mathrm{kg\,m^{-3}}$\\
$\sigma$ & Stefan--Boltzmann constant & $\mathrm{W\,m^{-2}\,K^{-4}}$\\
$\sigma_{\rm c}$ & Molecular collisional cross section & $\mathrm{m^2}$\\
$\tau_{\imath}$ & Phase change rate & $\mathrm{kg\,m^{-3}\,s^{-1}}$\\
$\Phi_{\imath}$ & Radial gas diffusion rate of species $\imath$ & $\mathrm{kg\,m^{-2}\,s^{-1}}$\\
$\Psi_{\imath}$ & Latitudinal gas diffusion rate of species $\imath$ & $\mathrm{kg\,m^{-2}\,s^{-1}}$\\
$\psi$ & Porosity & --\\
$\omega$ & Angular velocity of nucleus rotation & $\mathrm{rad\,s^{-1}}$\\
\hline 
\hline
\end{tabular}
\caption{Functions and parameters with descriptions and units. Greek.}
\label{tab0b}
\end{center}
\end{table}

\subsection{Keplerian orbits} \label{sec_method_kepler}

Let $\mathbf{r}_{\rm s}$ be the position vector of a source region on the southern nucleus hemisphere, in an inertial frame with its $\hat{z}$ 
axis aligned with the nucleus positive spin pole, at the time of dust ejection $t=0$. Let $\mathbf{v}_{\rm s}$ be the dust ejection velocity vector in the same frame with one component due 
to gas drag acceleration along the local surface normal $\hat{n}_{\rm s}$ and another due to nucleus rotation,
\begin{equation} \label{eq:01}
\mathbf{v}_{\rm s}=v_{\rm d}\hat{n}_{\rm s}+\hat{z}\times\mathbf{r}_{\rm s}\omega
\end{equation}
where the angular velocity is $\omega=2\pi/P$ and $P$ is the nucleus rotational period. Then the normal to 
the dust chunk orbital plane is $\mathbf{h}=\mathbf{r}_{\rm s}\times\mathbf{v}_{\rm s}=\{h_{\rm x},\,h_{\rm y},\,h_{\rm z}\}$.\\

\noindent
Let $\mathbf{r}_{\rm t}'$ be the position vector of a target area on the northern nucleus hemisphere in the inertial frame at 
the dust ejection time $t=0$ (equivalently, the target area position vector in the body--fixed frame). At a later time the target area has position $\mathbf{r}_{\rm t}(t)=M_{\rm rot}(t)\mathbf{r}_{\rm t}'=\{r_{\rm tx},\,r_{\rm ty},\,r_{\rm tz}\}$ in the inertial frame because of nucleus rotation, where
\begin{equation} \label{eq:02}
M_{\rm rot}(t)=\left(\begin{array}{l}
\displaystyle \cos\omega t\hspace{0.2cm} -\sin\omega t\hspace{0.5cm} 0\\
\displaystyle \sin\omega t\hspace{0.65cm} \cos\omega t\hspace{0.5cm} 0\\
\displaystyle 0\hspace{1.5cm} 0\hspace{1.35cm} 1\\
\end{array}\right).
\end{equation}
The criterion for the target area being located within the dust orbital plane is then 
\begin{equation} \label{eq:03}
\mathbf{h}\cdot\mathbf{r}_{\rm t}(t_{\rm pl})=0,
\end{equation}
remembering that the orbital plane necessarily must pass through the center of mass of the nucleus that hosts the origin of the coordinate system. 
The crossing of the target area with the dust orbital plane will repeat twice during a nucleus rotation at instances of 
time $t_{\rm pl}$ obtained from Eq.~(\ref{eq:03}) and associated equations,
\begin{equation} \label{eq:04}
t_{\rm pl}=\frac{1}{\omega}\sin^{-1}\left(\sqrt{1-f^2}\right)
\end{equation}
where
\begin{equation} \label{eq:05}
\left\{\begin{array}{l}
\displaystyle f=-\frac{p}{2}\pm \sqrt{\frac{p^2}{4}-q}\\
\\
\displaystyle p=\frac{2ac}{a^2+b^2}\\
\\
\displaystyle q=\frac{c^2-b^2}{a^2+b^2}\\
\\
 \displaystyle a=h_{\rm x}r_{\rm tx}+h_{\rm y}r_{\rm ty}\\
\displaystyle b=-h_{\rm x}r_{\rm ty}+h_{\rm y}r_{\rm tx}\\
\displaystyle c=r_{\rm tz}h_{\rm z}
\end{array}\right..
\end{equation}
Solutions exist if $f$ is real and $f\in [-1,\,1]$ (i.e., the target area does cross the orbital plane).\\

\noindent
The $\{\mathbf{r}_{\rm s},\,\mathbf{v}_{\rm s}\}$ pair can be used to calculate orbital elements for the dust trajectory in ecliptic space, 
following standard methods (e.g., \shortciteNP{boulet91}). These orbital elements can be used to calculate the location $\mathbf{r}_{\rm d}(t_{\rm pl})$ of the 
dust chunk at $t_{\rm pl}$. The distance $d=|\mathbf{r}_{\rm d}(t_{\rm pl})-\mathbf{r}_{\rm t}(t_{\rm pl})|$ is a function of $v_{\rm d}$ only, for any given 
combination of source and target location. We calculate $d$ as function of $v_{\rm d}$, starting at $v_{\rm d}=0$ and incrementing in 
steps of $0.001\,\mathrm{m\,s^{-1}}$. Our condition for the existence of a Keplerian trajectory that connects the source and target regions is $\min(d)\leq 25\,\mathrm{m}$. 
It means that the dust trajectory crosses the circle swept up by the target area during nucleus rotation (co--location in space) and that the dust 
chunk and target area arrives to this interception point simultaneously (co--location in time).\\

\noindent
Co--location in space and time is a necessary but not sufficient criterion for airfall onto the target area. It is also required that the dust chunk 
does not intercept another part of the nucleus on its way to the target area. Therefore, all facets on the shape model that cross the dust trajectory orbital 
plane are located, their crossing times are determined and the dust chunk position at that time is calculated. If a co--location in 
space and time takes place prior to the dust chunk arrival time to the target area, the trajectory is discarded.\\

\noindent
The trajectory search was made by considering the SHAP5 version 1.5 shape model \shortcite{jordaetal16} of comet 67P degraded to 
$5\cdot 10^4$ facets. Out of the $\sim 2.5\cdot 10^4$ facets on the southern hemisphere treated as potential source regions, there were 
$151$ facets that had the correct location and orientation to enable successful trajectories to the 31 northern hemisphere sites.

\subsection{Realistic gravity and numerical orbit integration} \label{sec_method_numorb}

The method in Sec.~\ref{sec_method_kepler} treats the nucleus as a point mass. That allows us to consider all $\sim 400$ million 
combinations of source regions ($\sim 2.5\cdot 10^4$), $v_{\rm d}$ values ($\sim 500$), and target areas ($\sim 30$) at a 
relatively low computational cost because the approach is essentially analytical. However, the nucleus is irregular and the gravity 
field deviates from that of a point source. We evaluate the error introduced by our point--mass assumption by comparing 
the Keplerian orbits with trajectories obtained by numerical integration of the equation of motion in a realistic gravity field.\\

\noindent
The nucleus gravitational potential $U$ at the center of each nucleus facet, the gravitational force vector $\mathbf{F}_{\rm g}(x,\,y,\,z)=\nabla U$ at 
any location exterior to the body surface, and the Laplacian $\nabla^2U$ (taking the value 0 outside the surface of the nucleus and $-4\pi$ within it) were 
calculated using the method of \shortciteN{wernerscheeres97}. This method applies to bodies with an arbitrarily complex shape described 
by a surface represented by polyhedrons under the assumption of constant density in the interior. Validation of the implementation was made by verifying 
that $\nabla U$ conformed with the Newtonian solution for polyhedrons forming a sphere, and that $U$ calculated for the highly 
irregular shape of comet 67P reproduces the potential map of \shortciteN{kelleretal17} in their Fig.~4. In order to represent the shape of 
67P we rely on the SHAP5 version 1.5 shape model \shortcite{jordaetal16} degraded to $5\cdot 10^3$ facets.\\

\noindent
In order to solve the equation of motion numerically an $8^{\rm th}$ order Gauss--Jackson integrator was implemented (\shortciteNP{fox84}; 
\shortciteNP{berryhealy04}) using a $4^{\rm th}$ order Runge--Kutta method \shortcite{burdenfairs93} as a startup formula. In order to validate 
the implementation the orbit of Comet C/1995~O1 (Hale--Bopp) was integrated from January 1993 through 2010 including the perturbations 
of the eight planets. This included a $0.77\,\mathrm{AU}$ encounter with Jupiter on April 5, 1996 that drastically modified the orbit, e.g., 
changed the orbital period from $\mathcal{P}=4221\,\mathrm{yr}$ to $2372\,\mathrm{yr}$. The solution was compared to that obtained with 
the RMVS3 version of the integrator \texttt{SWIFT} \shortcite{levisonduncan94} and the final positions of the comet according to the 
two codes differed by merely $70\,\mathrm{m}$.

\subsection{Nucleus thermophysics} \label{sec_method_nucthermo}

\noindent
The search for uninterrupted dust trajectories described in Sec.~\ref{sec_method_kepler} identified 151 unique source 
facets on the southern hemisphere responsible for airfall at the 31 target regions. The purpose of the thermophysical nucleus modeling is to calculate the 
local water production rates for those source regions as a function of nucleus rotational phase during a 315 day period from the inbound equinox on 
May 11, 2015, throughout the perihelion passage up to the outbound equinox on March 21, 2016. At this time the southern hemisphere was illuminated and the airfall in the 
north would have peaked. We assume that all regions have equal potential for activity when subjected to identical illumination sequences. This is consistent 
with the surface $\mathrm{H_2O}$ activity distribution of \shortciteN{fougereetal16b} that shows little variability within the southern hemisphere.\\

\noindent
All facets on the shape model that have an unobstructed view of at least one source facet were identified (here referred to as ``surrounding terrain'').  The $\sim 3\cdot 10^4$ facets in 
surrounding terrain have the capacity of shadowing the source facets by temporarily switching off the direct solar flux, and radiation that is scattered in the visual or emitted 
in the infrared from these facets will illuminate the sources (such nucleus self heating elevates illumination levels at all times and are particularly 
important when the sources are in shadow or experiencing night--time). We used the model presented by \shortciteN{davidssonandrickman14} in 
order to calculate the illumination conditions, including shadowing and self heating. For a given source facet $j$ we evaluate whether the Sun is fully visible 
($v_{j\odot}=1$) or if the facet is partially shadowed (rounded to $v_{j\odot}=1/3$ or $v_{j\odot}=2/3$) or fully shadowed by topography ($v_{j\odot}=0$) in $10^{\circ}$ 
increments of nucleus rotational phase throughout the time period, applying the spin axis in equatorial right ascension and declination $\{\alpha,\,\delta\}=\{69.54^{\circ},\,64.11^{\circ}\}$ 
of 67P \shortcite{preuskeretal15} and taking the time--dependent nucleus rotation period into account (e.g., \shortciteNP{kelleretal15b}).  For simplicity we assume that the 
combined scattering and emission of a surrounding terrain facet $k$ equals the local incident direct illumination (i.e., there is no loss or gain because of heat conduction from or to the surface). 
The fraction of the radiation emanating from $k$ that reaches a source facet $j$ is given by the view factor (e.g., \shortciteNP{ozisik85}) 
\begin{equation} \label{eq:08}
\mathcal{F}_{jk}=\frac{s_k\cos\beta_j\cos\beta_k}{\pi r_{jk}^2}.
\end{equation}
Here, $s_k$ is the surface area of facet $k$, the angles between the surface normals of $j$ and $k$ and their connection line are $\beta_j$ 
and $\beta_k$, respectively, and $r_{jk}$ is the distance between the two facets. For each source facet $j$ we solve the energy conservation equation
\begin{equation} \label{eq:06}
\rho c_{\rm s}\frac{\partial T_j(x,t)}{\partial t}=\frac{\partial}{\partial x_j}\left(\kappa\frac{\partial T_j(x,t)}{\partial x_j}\right)
\end{equation}
for the temperature $T_j$ by using the Finite Element Method. The surface boundary condition of Eq.~(\ref{eq:06}) is given by
\begin{equation} \label{eq:07}
\begin{array}{c}
\displaystyle  \frac{S_{\odot}v_{j\odot}(1-A)\mu_j(t)}{r_{\rm h}^2}=\varepsilon\sigma T_j(0,t)^4- (1-A)\sum_{k\not=j}\mathcal{F}_{jk}\left(\frac{S_{\odot}v_{k\odot}A\mu_k(t)}{r_{\rm h}^2}+\varepsilon\sigma T_k(0,t)^4 \right)\\
\displaystyle \\
\displaystyle +f_i\alpha_sZ(T_j)\mathcal{L}-\kappa\frac{\partial T}{\partial x_j}\Big|_{x_j=0}.\\
\end{array}
\end{equation}
From left to right, the terms in Eq.~(\ref{eq:07}) denote the absorbed flux from direct solar illumination; thermally emitted infrared radiation;  self heating in the form of scattered optical and 
thermally emitted radiation from other facets; energy consumption by sublimation of near--surface ice; and heat conducted from the surface to the interior or vice versa depending on 
the sign of the temperature gradient. The boundary condition of Eq.~(\ref{eq:06}) at the bottom of the calculational domain (typically placed at ten times the diurnal thermal skin depth below the surface) is a 
vanishing temperature gradient;
\begin{equation} \label{eq:09}
\frac{\partial T_j}{\partial x_j}\Big|_{x=x_{\rm max}}=0.
\end{equation}
Solving Eqs.~(\ref{eq:06}) with its boundary conditions yields $T_j=T_j(x,t)$ and the vapor production rate then is calculated as 
\begin{equation} \label{eq:10}
Q_j=f_i\alpha_sZ(T_j(0,t))
\end{equation}
where $f_i$ is the volume fraction of water ice, $\alpha_{\rm s}$ is the sublimation coefficient \shortcite{kossackietal99b},
\begin{equation} \label{eq:11}
\alpha_{\rm s}=1-\frac{1}{a_1}+\frac{1}{a_1}\tanh\left(-a_3\tan\left(\pi\frac{T-a_2}{273-a_2}-\frac{\pi}{2}\right)\right),
\end{equation}
where $a_1=2.342$, $a_2=150.5$, and $a_3=4.353$, and the Hertz--Knudsen formula is
\begin{equation} \label{eq:12}
Z=p_{\rm sat}(T)\sqrt{\frac{m}{2\pi k_{\rm B}T}}
\end{equation}
where the saturation pressure of water vapor (in $\mathrm{Pa}$) is given by \shortciteN{fanaleandsalvail84}
\begin{equation} \label{eq:13}
p_{\rm sat}(T)=3.56\times 10^{12}\exp\left(-\frac{6141.667}{T}\right).
\end{equation}
%

\subsection{Coma structure} \label{sec_method_coma}

The purpose of this Section is to present the methods used for calculating the number density $n(h)$, translational temperature $T(h)$, and 
drift (expansion) speed $W(h)$ as functions of height $h$ above the nucleus surface (along the local surface normal). This calculation is non--trivial because outgassing 
from a solid surface into vacuum yields a vapor that initially is deviating drastically from thermodynamic equilibrium (e.g., \shortciteNP{cercignani00}), 
i.e., the velocity distribution function does not resemble the Maxwell--Boltzmann function, and hydrodynamic relations as such do not apply because 
they are based on a simplified Boltzmann equation where the collision integral is zero. Molecular collisions will gradually evolve the gas toward 
thermodynamic equilibrium, but only if molecular collisions are common. Therefore, there is a kinetic region known as the Knudsen layer above 
the nucleus surface that may transit into a hydrodynamic coma at some height, unless the Knudsen layer extends to infinity. We need to deal with 
both cases, treating the first possibility (strong sublimation) in Sec.~\ref{sec_method_coma_strong} and the second (weak sublimation) in Sec.~\ref{sec_method_coma_weak}. 
For various discussions about cometary Knudsen layers see, e.g., \shortciteNP{skorovandrickman98}, \citeyearNP{skorovandrickman99}; \shortciteNP{huebnerandmarkiewicz00}; 
\shortciteNP{crifoetal02}; \shortciteNP{davidssonandskorov04}; \shortciteNP{davidsson08}; \shortciteNP{davidssonetal10}.\\

\noindent
In the discussion that follows, we use the following subscripts to distinguish various regions; a sub--surface region responsible for feeding the coma ``$\mathrm{s}$''; 
the bottom of the Knudsen layer ``0''; the top of the Knudsen layer ``1''; a location within the hydrodynamic part of the coma ``2''; and infinity ``$\infty$''. We 
start by defining the difference between strong and weak sublimation.\\ 

\noindent
At a given moment the nucleus outgassing rate and surface temperature of a given facet are given by $\{Q_{\rm s},\,T_{\rm s}\}$. 
The sub--surface reservoir number density capable of sustaining the production rate $Q_{\rm s}$ is
\begin{equation} \label{eq:14}
n_{\rm s}=Q_{\rm s}\sqrt{\frac{2\pi}{mk_{\rm B}T_{\rm s}}}
\end{equation}
and the Knudsen layer bottom number density is 
\begin{equation} \label{eq:15}
n_0=\frac{1}{2}(1-\alpha_{\rm bf})n_{\rm s}
\end{equation}
where $\alpha_{\rm bf}$ is the surface backflux fraction (i.e., the fraction of molecules that return to the surface). The molecular mean free path is
\begin{equation} \label{eq:16}
\lambda=\frac{1}{\sigma_{\rm c} n_0}
\end{equation}
where $\sigma_{\rm c}$ is the molecular collisional cross section. The Knudsen layer is taken to have a thickness of $K_{\rm t}=30\lambda$ 
(90--99\% of thermodynamic equilibrium is achieved at 20--$200\,\lambda$ according to \shortciteNP{cercignani00}). For 
a nucleus curvature radius $R_{\rm n}$ we define strong sublimation as $R_{\rm n}/(R_{\rm n}+30\lambda)>0.9$, i.e., the 
Knudsen layer is relatively thin compared to the nucleus radius and the degree of gas expansion within the Knudsen layer is small. 
Such conditions are typical on the dayside at small heliocentric distances. We define weak sublimation as $R_{\rm n}/(R_{\rm n}+30\lambda)\leq 0.9$, i.e., 
the gas expansion is substantial within the Knudsen layer and downstream hydrodynamic conditions may not be reached. Such conditions 
are typical at large heliocentric distances, and on the nightside at any point in the orbit.

\subsubsection{Strong sublimation} \label{sec_method_coma_strong}

We first describe our applied solution for the Knudsen layer, and then for the downstream hydrodynamic coma.  Classical gas kinetic theory 
(\shortciteNP{anisimov68}; \shortciteNP{ytrehus77}) shows that the temperature and number density at the 
top of the Knudsen layer are related to the sub--surface parameters (the so--called ``jump conditions'') through
\begin{equation} \label{eq:17}
\left\{\begin{array}{c}
\displaystyle \frac{T_1}{T_{\rm s}}=\left(-\frac{1}{8}\sqrt{\pi}S_1+\sqrt{1+\frac{\pi}{64}S_1^2}\right)^2\\
\\
\displaystyle \frac{n_1}{n_{\rm s}}=\frac{F+\sqrt{\frac{T_1}{T_{\rm s}}}G}{2\exp(-S_1^2)}
\end{array}\right.
\end{equation}
where 
\begin{equation} \label{eq:18}
\left\{\begin{array}{c}
\displaystyle F=-\sqrt{\pi}S_1{\rm erfc}(S_1)+\exp(-S_1^2)\\
\\
\displaystyle G=(2S_1^2+1){\rm erfc}(S_1)-\frac{2}{\sqrt{\pi}}\exp(-S_1^2)
\end{array}\right.
\end{equation}
where $\mathrm{erfc}()$ is the complimentary error function and the speed ratio is $S_1=\sqrt{5/6}M_1$ where 
the Mach number is
\begin{equation} \label{eq:19}
M=W\sqrt{\frac{m}{\gamma k_{\rm B}T}}.
\end{equation}
Here, the heat capacity ratio $\gamma$ and the number of rotational degrees of freedom $\zeta$ are related by
\begin{equation} \label{eq:20}
\gamma=\frac{\zeta+5}{\zeta+3}.
\end{equation}
The ratio of the number of molecules traveling back towards the surface at the bottom relative to the top of 
the Knudsen layer is 
\begin{equation} \label{eq:21}
\beta^-=\frac{2(2S_1^2+1)\sqrt{\frac{T_1}{T_{\rm s}}}-2\sqrt{\pi}S_1}{F+\sqrt{\frac{T_1}{T_{\rm s}}}G}.
\end{equation}
Furthermore, the analysis showed that the Mach number approached unity at the top of the Knudsen layer ($M_1\approx 1$), 
which means that $T_1/T_{\rm s}\approx 0.67$, $n_1/n_{\rm s}\approx 0.21$ and that $\beta^-$ implies $\alpha_{\rm bf}\approx 0.18$ (see Eq.~\ref{eq:15}). 
However, these results apply for monatomic vapor and are not directly applicable for water molecules that have $\zeta=3$ and $\gamma=4/3$. 
Assuming that $T_1/T_{\rm s}\approx 0.67$ still holds, requiring that $M_1=1$ for $\gamma=4/3$, recognizing that mass conservation 
yields $W_1=Q_{\rm s}/mn_1$ on top of the Knudsen layer while applying Eqs.~(\ref{eq:14}) and (\ref{eq:19}) yields the following 
jump condition for water vapor,
\begin{equation} \label{eq:22}
\left\{\begin{array}{c}
\displaystyle \frac{T_1}{T_{\rm s}}=0.67\\
\\
\displaystyle \frac{n_1}{n_{\rm s}}=\sqrt{\frac{T_{\rm s}}{2\pi\gamma T_1}}\approx 0.42.
\end{array}\right.
\end{equation}
Using the ratios in Eq.~(\ref{eq:22}) we apply
\begin{equation} \label{eq:23}
\left\{\begin{array}{l}
\displaystyle n(h)=\frac{n_1-n_0}{K_{\rm t}}h+n_0
\\
\\
\displaystyle W(h)=\frac{Q_{\rm s}}{mn(h)}\\
\\
\displaystyle T(h)=T_1\\
\end{array}\right.
\end{equation}
for heights $0\leq h\leq K_{\rm t}$ above the nucleus surface, i.e., we assume a linear reduction of the number density from $n_0$ 
near the surface (Eq.~\ref{eq:15}) to $n_1$ at $K_{\rm t}$, evaluate the drift speed according to mass conservation, and assume 
that the translational temperature remains quasi--constant throughout the Knudsen layer. The last property is motivated by numerical 
solutions to the Boltzmann equation obtained with the Direct Simulation Monte Carlo (DSMC) technique, see \shortciteN{davidsson08}.\\

\noindent
For $h>K_{\rm t}$ we apply an analytical hydrodynamic solution valid for isentropic expansion summarized by \shortciteN{davidssonetal10}, 
see their Eq~(1)--(6), that merges smoothly with the Knudsen layer solution. Specifically, we first find the Mach number $M_2$ at height $h$ by numerically solving
\begin{equation} \label{eq:24}
\left(\frac{h+R_{\rm n}}{K_{\rm t}+R_{\rm n}}\right)^2=\frac{1}{M_2}\left(\frac{1+\frac{1}{2}(\gamma-1)}{1+\frac{1}{2}(\gamma-1)M_2^2}\right)^{-\frac{\gamma+1}{2(\gamma-1)}}.
\end{equation}
Next, the constant $\mathcal{C}$ given by 
\begin{equation} \label{eq:25}
\mathcal{C}=Tn^{1-\gamma}
\end{equation}
is determined by using the known $\{n_1,\,T_1\}$ values and the parameter $\mathcal{D}(h)$ is evaluated which yields the product of 
number density and drift speed at the corresponding height from mass conservation,
\begin{equation} \label{eq:26}
\mathcal{D}=\frac{Q_{\rm s}}{m}\left(\frac{R_{\rm n}}{R_{\rm n}+h}\right)^2=nW.
\end{equation}
Equations~(\ref{eq:25})--(\ref{eq:26}) are inserted into Eq.~(\ref{eq:19}) which is solved for $n$, thereby allowing that parameter 
to be determined,
\begin{equation} \label{eq:27}
n(h)=\left(\frac{M_2}{\mathcal{D}}\sqrt{\frac{\gamma k_{\rm B}\mathcal{C}}{m}}\right)^{\left[\frac{1}{2}(1-\gamma)-1\right]^{-1}}.
\end{equation}
With $n_2=n(h)$ known, $W_2=W(h)$ follows from Eq.~(\ref{eq:26}) and $T_2=T(h)$ from  Eq.~(\ref{eq:25}), thus the gas parameters for 
the strong sublimation case can be evaluated.

\subsubsection{Weak sublimation} \label{sec_method_coma_weak}

When collisions are rare, the gas is unable to achieve thermodynamic equilibrium. An accurate description of the coma in 
this case requires numerical solutions to the Boltzmann equation. Such sophisticated treatment of the problem is beyond the 
scope of this paper and we apply a simple parameterized description of the coma that reproduce general characteristics 
of DSMC solutions reasonable well (see Sec.~\ref{sec_results}).\\

\noindent
The average kinetic energy of a water molecule upon release from the nucleus is 
\begin{equation} \label{eq:28}
E_{\rm t}=\frac{3}{2}k_{\rm B}T_{\rm s}
\end{equation}
and its average internal energy because of molecular rotation is
\begin{equation} \label{eq:28b}
E_{\rm r}=\frac{\zeta}{2}k_{\rm B}T_{\rm s}
\end{equation}
(see, e.g., \shortciteNP{bird94}). If collisions are very common ($\lambda\ll R_{\rm n}$) the far downstream molecular 
velocity vectors are radially aligned (i.e., essentially parallel to each other) and all speed variability has been removed through 
energy equipartitioning. Because there are no random velocities about the local mean, $T_{\infty,\,\lambda\ll R_{\rm n}}=0$ and all kinetic energy is in bulk drift. 
Therefore, an ideal hydrodynamic gas that expands towards infinity ($n\rightarrow 0$)  has $T\rightarrow 0$  (so that Eq.~\ref{eq:25} remains 
valid at arbitrarily low $n$). Because expansion leads to translational temperature cooling, and because collisions are common, energy is transferred from 
the internal to the kinetic mode and the rotational temperature cools to $T_{\rm r}=0$ at infinity as well. Thus, all available energy is in bulk drift kinetic energy 
$mW_{\infty,\,\lambda\ll R_{\rm n}}^2/2=E_{\rm t}+E_{\rm r}$ and therefore
\begin{equation} \label{eq:29}
W_{\infty,\,\lambda\ll R_{\rm n}}=\sqrt{\frac{(3+\zeta)k_{\rm B}T_{\rm s}}{m}}.
\end{equation}
If collisions are very few ($\lambda\gg R_{\rm n}$) the downstream velocity vectors are still radially aligned purely because of geometric reasons. However, a 
molecular speed dispersion remains that collisions have not been able to remove. Therefore, the downstream flow is characterized by a non--zero translational \emph{freeze--out} 
temperature $T_{\infty,\,\lambda\gg R_{\rm n}}=T_{\rm fr}\not=0$. Because most translational energy still is in bulk drift kinetic energy ($T_{\rm fr}$ is low) 
and because very little rotational energy has been transferred ($T_{\rm r}$ remains quasi--constant) the bulk drift kinetic energy is $mW_{\infty,\,\lambda\gg R_{\rm n}}^2/2=E_{\rm t}$ and therefore

\begin{equation} \label{eq:30}
W_{\infty,\,\lambda\gg R_{\rm n}}=\sqrt{\frac{3k_{\rm B}T_{\rm s}}{m}}.
\end{equation}
For the transition regime $0<\lambda<R_{\rm n}$ we assume that there is a linear transition from $\{T_{\infty},\,W_{\infty}\}_{\lambda\ll R_{\rm n}}$ to $\{T_{\infty},\,W_{\infty}\}_{\lambda\gg R_{\rm n}}$. 
The temperature and drift speed change with height, starting at their near--surface values $\{T_0,\,W_0\}$ and gradually approaching their downstream values $\{T_{\infty},\,W_{\infty}\}$. 
The rate of change depends on the number of molecular collisions that take place per time unit. That number depends on the local gas density, here assumed to follow  $n\propto h^{-2}$.
We realize these assumptions through the expressions
\begin{equation} \label{eq:31}
\left\{\begin{array}{l}
\displaystyle W(h)=\left[1-\left(\frac{R_{\rm n}}{R_{\rm n}+h}\right)^2\right]\left(\sqrt{\frac{(3+\zeta\max\{1-\lambda_0/R_{\rm n},\,0\})k_{\rm B}T_{\rm s}}{m}}-W_0\right)+W_0\\
\\
\displaystyle T(h)=\left[1-\left(\frac{R_{\rm n}}{R_{\rm n}+h}\right)^2\right]\Big(T_{\rm fr}-T_{\rm fr}\max\{1-\lambda_0/R_{\rm n},\,0\}-T_0\Big)+T_0\\
\\
\displaystyle n(h)=\frac{\mathcal{D}(h)}{W(h)}
\end{array}\right.
\end{equation}
where $\lambda_0=(n_0\sigma)^{-1}$ and $\mathcal{D}(h)$ is evaluated as in Eq.~(\ref{eq:26}). Inspired by the numerical DSMC solutions to the 
Boltzmann equation by \shortciteN{davidssonetal10} we set $T_{\rm fr}=20\,\mathrm{K}$. The expression for $W(h)$ in Eq.~(\ref{eq:31}) can be 
understood as follows. The first term within the curved bracket is the downstream terminal drift speed. It is close to Eq.~(\ref{eq:29}) when $\lambda_0\approx 0$ and 
falls linearly with $\lambda_0$ towards Eq.~(\ref{eq:30}) as $\lambda_0\rightarrow R_{\rm n}$, and remains at that value for even more diluted gas ($\lambda_0\geq R_{\rm n}$) 
because of the maximum function. The expression within the curved bracket therefore measures the largest considered increase above $W_0$, the near--surface value. The 
expression within the square bracket (confined to the $[0,1]$ interval) regulates how quickly $W(h)$ grows from $W_0$ at $h=0$ to $W_{\infty}$ as $h\rightarrow\infty$. It is an inverse--square law of height, 
because of the previously mentioned relations between collision frequency, rate of change of the kinetic properties, and number density. The expression for $T(h)$ follows a similar logic.

\subsection{Dust ejection} \label{sec_method_eject}

In order to calculate the velocity reached by a dust chunk of a particular size we need to solve the equation of motion along the local 
surface normal,
\begin{equation} \label{eq:32}
m_{\rm d}\frac{d^2\,h}{d\,t^2}=F_{\rm g}(h)+F_{\rm drag}(h,u_{\rm d})
\end{equation}
where the chunk mass is $m_{\rm d}$ and the chunk speed is $u_{\rm d}=dh/dt$. We assume that the initial velocity is zero, i.e., that the mechanism that 
liberates dust particles from their attachment to the nucleus does not transfer momentum to the dust. For the current application we do not need to 
specify the mechanism that is responsible for the chunk release. However, we note that in laboratory experiments \shortciteN{ratkekochan89} observed oscillation of dust 
aggregates at a frequency of 1--$100\,\mathrm{Hz}$ that lasted for up to 10 minutes before their final release. This suggests that gas drag ultimately is 
responsible not only for the acceleration of dust but also for its liberation, through a fatigue process that gradually reduces the initially strong cohesion of the 
granular nucleus material to the point the chunk breaks free. We evaluate the local gravitational force $F_{\rm g}(h)$ (directed towards the nucleus) at height $h$ by using the \shortciteN{wernerscheeres97} formalism discussed 
in Sec.~\ref{sec_method_numorb}.  Therefore, the point--mass assumption is used to determine the orbital speed $v_{\rm d}$ that is necessary to reach the target site, 
but realistic gravity is used when calculating the initial acceleration from rest to the point that $v_{\rm d}$ is reached. This approach is reasonable, given that realistic gravity rapidly 
approaches the point--mass solution with height. We evaluate the gas drag force $F_{\rm drag}(h,\,u_{\rm d})$ 
by utilizing the local coma solution $\{n(h),\,T(h),\,W(h)\}$ obtained in Sec.~\ref{sec_method_coma} for chunks of cross section $\mathcal{A}$ and applying \ (e.g., \shortciteNP{gombosi94}),
\begin{equation} \label{eq:33}
\left\{\begin{array}{c}
\displaystyle F_{\rm drag}(h,\,u_{\rm d})=\frac{1}{2}\mathcal{A} C_{\rm D}(h,\,u_{\rm d}) n(h)\left(W(h)-u_{\rm d}(h)\right)^2\\
\\
\displaystyle C_{\rm D}(h,\,u_{\rm d})=\frac{\left(2\mathcal{W}^2+1\right)\exp\left(-\mathcal{W}^2\right)}{\sqrt{\pi}\mathcal{W}^3}+\frac{\left(4\mathcal{W}^4+4\mathcal{W}^2-1\right){\rm erf}\left(\mathcal{W}\right)}{2\mathcal{W}^4}.
\end{array}\right.
\end{equation}
Here, $\mathrm{erf()}$ is the error function and $\mathcal{W}$ is given by
\begin{equation} \label{eq:34}
\mathcal{W}=\left(W(h)-u_{\rm d}(h)\right)\sqrt{\frac{m}{2k_{\rm B}T(h)}}.
\end{equation}
Equation~(\ref{eq:32}) is solved numerically for every source region, for every $10^{\circ}$ advancement of the nucleus rotational phase during the considered $\sim 11\,\mathrm{month}$ 
period around the perihelion passage, for 19 chunk size classes (with dimensions that are $0.05\mathcal{A}_{\rm crit}$--$0.95\mathcal{A}_{\rm crit}$ in increments of $0.05\mathcal{A}_{\rm crit}$, 
where $\mathcal{A}_{\rm crit}$ is the critical maximum liftable chunk cross section for which $F_{\rm g(0)}+F_{\rm drag}(0,\,0)=0$ is valid at the considered location and time).  We here consider 
chunks shaped as oblate ellipsoids with axis ratio 4, exposing their largest cross section to the gas flow at $h=0$ ($\sim 2.5$ times that of an equal--mass sphere), that turn with constant radial velocity to expose their smallest cross section ($\sim 0.63$ times that of an equal--mass sphere) during the first $500\,\mathrm{m}$ of flight. The chunk shape is close to the axis ratio of 5 needed in order 
to match the particle speeds measured by Rosetta/GIADA according to \shortciteN{ivanovskietal17a}. The time needed for the chunks to reach the height of $500\,\mathrm{m}$ is similar to that 
needed by the aerodynamic torque to set non--spherical chunks in rotation according to simulations by \shortciteN{ivanovskietal17b}. The assumed decrease of the chunk cross section with time shortens the time needed for the particle to reach its terminal velocity, and makes $F_{\rm drag}$ diminish in strength faster than $F_{\rm g}$, thereby illustrating how the assumed dominance of nucleus gravity over drag (after the initial acceleration) can be motivated. We select the chunk size $a_{\rm crit}(t)$ that reaches a quasi--terminal velocity of $u_{\rm d,\,max}$ that is as close to $v_{\rm d}$ as possible. This is considered  the best representative of the chunk type that feed a particular target area at a given time. Note that this size changes continuously during nucleus rotation and orbital motion.

\subsection{Airfall} \label{sec_method_airfall}

In order to estimate the amount of mass that is channeled toward a certain target region it is necessary to know the total amount of dust that 
is being ejected from a source region at a given time, and the fraction of that mass carried by chunks of size $a_{\rm crit}(t)$. The total dust 
mass flux is taken as $Q_{\rm d}=4Q_{\rm s}$ \shortcite{rotundietal15}. The estimate of the fraction of this mass carried by $a_{\rm crit}(t)$--chunks 
requires that we specify a dust size distribution function.\\

\noindent
For $3.7\,\mathrm{mm}\leq a_{\rm crit}\leq 1.62\,\mathrm{cm}$ we use the size distribution for chunks observed in Philae/CIVA images at Abydos, 
consisting of three segments with different slopes \shortcite{pouletetal17}. We extrapolate the $\leq 5\,\mathrm{mm}$ part of this distribution to 
smaller sizes. For $4\,\mathrm{cm}\leq a_{\rm crit}\leq 1\,\mathrm{m}$ we use the size distribution for chunks observed in Philae/ROLIS images 
at Agilkia \shortcite{mottolaetal15}. We use this distribution to bridge the $1.62$--$4\,\mathrm{cm}$ gap in the measured data, and to extrapolate 
to larger sizes. With a differential size distribution ($dN/dD\propto D^{-q_{\rm s}}$) slope of $q_{\rm s}=3.8$ \shortcite{mottolaetal15} truncated at $D_{\rm min}=0.0162\,\mathrm{m}$ and 
$D_{\rm max}=1\,\mathrm{m}$ the amount of mass at diameters $D\geq D_{\rm c}=0.1\,\mathrm{m}$ is given by 
\begin{equation} \label{eq:35}
\mathcal{F}=\frac{\int_{D_{\rm c}}^{D_{\rm max}}D^{3-q_{\rm s}}\,dD}{\int_{D_{\rm min}}^{D_{\rm max}}D^{3-q_{\rm s}}\,dD}=\frac{D_{\rm max}^{4-q_{\rm s}}-D_{\rm c}^{4-q_{\rm s}}}{D_{\rm max}^{4-q_{\rm s}}-D_{\rm min}^{4-q_{\rm s}}}
\end{equation}
and amounts to $\mathcal{F}=66\%$. If this slope is extrapolated to $D_{\rm min}=1\,\mathrm{mm}$ then $\mathcal{F}=80\%$ for $D_{\rm c}=1\,\mathrm{cm}$ and 
$\mathcal{F}=49\%$ for $D_{\rm c}=0.1\,\mathrm{m}$. It is clear that a majority of the mass is carried by chunks that are cm--sized or larger.\\


\noindent
We truncate the size distribution function at the maximum ejectable size at a given source location and time, normalize the size distribution so that 
it carries the entire mass being ejected, and determine the fraction of mass carried by chunks with 
sizes in the range $a_{\rm crit}(t)/2$--$2a_{\rm crit}(t)$.\\

\noindent
The total airfall on a given target area is calculated by integrating the mass over all contributing source regions and time. We consider the 
activity of most target areas to be so low during the mass accumulation period that we assume the airfall has free access to the surface. The 
exception is target areas \#30 and \#31 (see Table~\ref{tab1}) that themselves are located on the southern hemisphere and are strongly illuminated at times. We 
therefore calculate the maximum liftable size in those locations versus time and only accept airfall chunks that are larger (i.e., we assume 
that the smaller chunks will be re--ejected within a short period of time if they manage to reach the surface in the first place). Therefore, sites \#30 and \#31 in Imhotep 
are partially self--cleaning. The thickness of the accumulated layer at each site is calculated from the incident mass flux assuming that 
chunks build a deposit with the same bulk density ($\rho=535\,\mathrm{kg\,m^{-3}}$) as the nucleus \shortcite{preuskeretal15}.

\subsection{Thermophysics of coma chunks} \label{sec_method_grainthermo}

During their flight through the coma the airfall chunks will lose some of the ice they carry. In order to calculate the typical volatile 
loss from chunks and to estimate the ice abundance in fresh airfall deposits we apply the code \texttt{NIMBUS} (Numerical Icy Minor Body evolUtion Simulator) 
developed by \shortciteN{davidsson20}. \texttt{NIMBUS} considers a rotating spherical body consisting of a porous mixture of dust and ice, 
and tracks the internal ice sublimation, vapor condensation, and diffusion of gas and heat in the radial and latitudinal directions over time during body rotation. 
We refer to \shortciteN{davidsson20} for a detailed description of the model but here state and briefly discuss the governing equations. 
The energy conservation equation is given by
\begin{equation} \label{eq:36}
\begin{array}{c}
\displaystyle \rho c(T)\frac{\partial T}{\partial t}=\frac{1}{r^2}\frac{\partial}{\partial r}\left(\kappa(\psi,\,T)r^2\frac{\partial T}{\partial r}\right)+\frac{1}{r\sin l}\frac{\partial }{\partial l}\left(\kappa (\psi,\,T)\frac{\sin l}{r}\frac{\partial T}{\partial l}\right)\\
\\
\displaystyle  -\sum_{\imath=4}^{n_{\rm sp}}q_{\imath}(p_{\imath},\,T)\mathcal{L}_{\imath}(T)+\sum_{\imath=2}^{n_{\rm sp}}\sum_{\jmath=5}^{n_{\rm sp}}\left( q'_{\imath}(T)\left\{H_{\imath}-F_{\imath\jmath}\mathcal{L}_{\jmath}(T)\right\}\right)-\sum_{\imath=4}^{n_{\rm sp}}g_{\imath}\left(\Phi_{\imath}\frac{\partial T}{\partial r}-\frac{\Psi_{\imath}}{r}\frac{\partial T}{\partial l}\right)+R_*\\
\end{array}
\end{equation}
with the following terms going left to right; 1) change of the internal energy; 2) radial heat conduction; 3) latitudinal heat conduction; 4) sublimation of ice and recondensation 
of vapor; 5) energy release during crystallization of amorphous water ice and energy consumption during its release of occluded $\mathrm{CO}$ and $\mathrm{CO_2}$, as well as 
energy consumption during $\mathrm{CO}$ segregation from its partial entrapment in condensed  $\mathrm{CO_2}$; 6) advection during radial and latitudinal gas diffusion; 7) heating 
by radioactive decay. The indices in Eq.~(\ref{eq:36}) refer to refractories ($\imath=1$), amorphous water ice ($\imath=2$), cubic water ice ($\imath=3$), hexagonal (crystalline) water ice ($\imath=4$), 
carbon monoxide ($\imath=5$), and carbon dioxide ($\imath=6$). Species denoted by $\jmath$ are hosted within a more abundant and less volatile species denoted by $\imath$. The upper boundary condition to Eq.~(\ref{eq:36}) is given by
\begin{equation} \label{eq:37}
\frac{S_{\odot}(1-A)\mu(l,t)}{r_{\rm h}^2}=\sigma\varepsilon T_{\rm surf}^4-\kappa\frac{\partial T}{\partial r}\Big|_{r=R_{\rm n}},
\end{equation}
and balances absorbed solar radiation with net thermal radiation losses and heat conduction. The boundary condition at the body center is a vanishing temperature gradient. The mass 
conservation equation for vapor ($\imath\geq 4$) is given by 
\begin{equation} \label{eq:38}
\begin{array}{c}
\displaystyle \psi m_{\imath}\frac{\partial n_{\imath}}{\partial t}=-\frac{1}{r^2}\frac{\partial}{\partial r}\left(r^2\Phi_{\imath} \right)-\frac{1}{r\sin l}\frac{\partial }{\partial l}\left(\Psi_{\imath}\sin l\right)+q_{\imath}(p_{\imath},\,T)+\sum_{\jmath=2}^{n_{\rm sp}}F_{\jmath,\imath}q'_{\jmath}(p_{\jmath},\,T),
\end{array}
\end{equation}
with terms left to right; 1) changes to the gas density; 2) radial gas diffusion; 3) latitudinal gas diffusion; 4) release during sublimation and consumption during condensation; 
5) release of $\mathrm{CO}$ and $\mathrm{CO_2}$ during crystallization of amorphous water ice, and of $\mathrm{CO}$ from sublimating $\mathrm{CO_2}$ ice. The 
boundary conditions to Eq.~(\ref{eq:38}) consist of a quantification of the gas venting to space at the surface and a vanishing gas flux at the center. Finally, the mass conservation 
equation for ice ($i\geq 2$) is given by
\begin{equation} \label{eq:39}
\begin{array}{c}
\displaystyle \frac{\partial \rho_{\imath}}{\partial t}=-q_{\imath}(p_{\imath},\,T)+\tau_{\imath}(T),
\end{array}
\end{equation}
with terms left to right; 1) changes to the amount of ice; 2) changes due to sublimation of ice and recondensation of vapor; 3) changes due to other phase transitions (crystallization of amorphous 
water ice, transition from cubic to hexagonal water ice).\\

\noindent
In the current simulations we nominally only consider one volatile -- hexagonal (crystalline) water ice that initially constitutes $20\%$ of the body mass. 
However, one simulation also has condensed $\mathrm{CO_2}$ at a 5\% level relative to water by number. The specific heat capacity $c(T)$ and the heat conductivity 
$\kappa(\psi,\,T)$ are mass--weighted averages of temperature--dependent values measured in the laboratory for water ice (\shortciteNP{klinger80}, \citeyearNP{klinger81}), carbon 
dioxide (\shortciteNP{giauqueandegan37}; \shortciteNP{kuhrt84}) and 
refractories (forsterite from \shortciteNP{robieetal82} for $c$ and ordinary chondrites 
from \shortciteNP{yomogidamatsui83} for $\kappa$). The heat conductivity is corrected for porosity following \shortciteN{shoshanyetal02} and includes both solid state and 
radiative conduction. The volume mass production rates $q_{\imath}$ are standard expressions (e.g., \shortciteNP{mekleretal90}; \shortciteNP{prialnik92}; \shortciteNP{tancredietal94}) 
with $\mathrm{H_2O}$ and $\mathrm{CO_2}$ saturation pressures taken from \shortciteN{huebneretal06} and the gas diffusion fluxes ($\Phi_{\imath}$, $\Psi_{\imath}$) are evaluated as in 
\shortciteN{davidssonandskorov02b}. The radiogenic heating from a number of long--lived isotopes at chondritic 
abundances in the refractory component is included by default but has a completely negligible influence on the results in the current application.

\subsection{Longterm loss of volatiles} \label{sec_method_volatileloss}

In order to better understand the fate of water ice after it has been deposited as airfall we perform thermophysical simulations for 
a selection of target areas. These simulations rely on the same model as described in Sec.~\ref{sec_method_nucthermo}. The major 
difference is that the illumination conditions at the target areas (calculated for every $10^{\circ}$ of nucleus rotation) are not 
restricted to the orbital arc between the inbound and outbound equinoxes, but are made for the entire orbit.

\section{Results} \label{sec_results}

\subsection{The thickness of airfall debris layers} \label{sec_results_layerthickness}

\begin{figure}
\begin{center}
     \scalebox{0.6}{\includegraphics{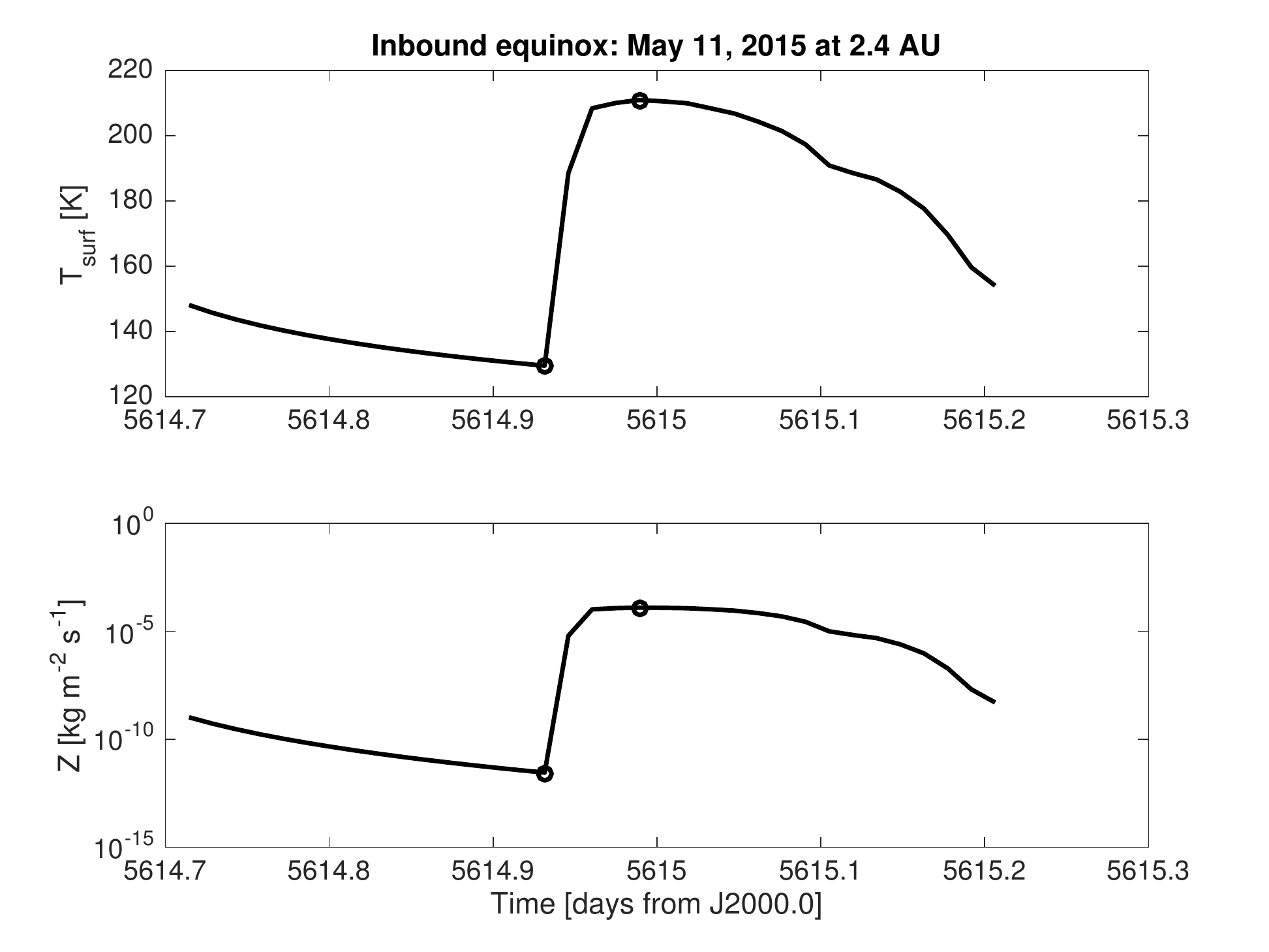}}
   \caption{Surface temperature (upper panel) and water production rate (lower panel) versus time for one of the source regions 
during a single nucleus rotation near inbound equinox. The circles denote minimum and maximum levels of activity.}
     \label{fig1}
\end{center}
\end{figure}

In order to exemplify the application of the models presented in Sec.~\ref{sec_method} we here describe sample results. The 
thermophysical modeling of the nucleus (Sec.~\ref{sec_method_nucthermo}) provides surface temperature $T_{\rm surf}(t)$ and water 
production rate $Q_{\rm s}$ during nucleus rotation. Figure~\ref{fig1} shows an example of diurnal temperature and production curves 
for a southern hemisphere source region\footnote{This randomly selected source is located in the Seth region at longitude $105.5^{\circ}\,\mathrm{W}$, 
latitude $21.1^{\circ}\,\mathrm{S}$, at a distance $1.33\,\mathrm{km}$ from the nucleus center (origin of the Cheops coordinate system, see  \shortciteNP{preuskeretal15})}. that contribute with material to target area \#3, at a 
time near the inbound equinox. In this particular case the surface temperature peaks near $T_{\rm surf}=211\,\mathrm{K}$ at local noon and falls to 
$T=129\,\mathrm{K}$ just prior to sunrise. The waviness of the temperature 
curve is a consequence of changes in shadowing and self heating conditions during nucleus rotation, in addition to the continuous 
changes of the solar height above the local horizon. The gas production rate is a strongly non--linear function of surface temperature 
(see Eqs.~\ref{eq:12}--\ref{eq:13}). Therefore, it varies by seven orders of magnitude at this particular location and time. This has 
strong implications for the properties of the near--nucleus coma.\\

\begin{figure}
\begin{center}
     \scalebox{0.6}{\includegraphics{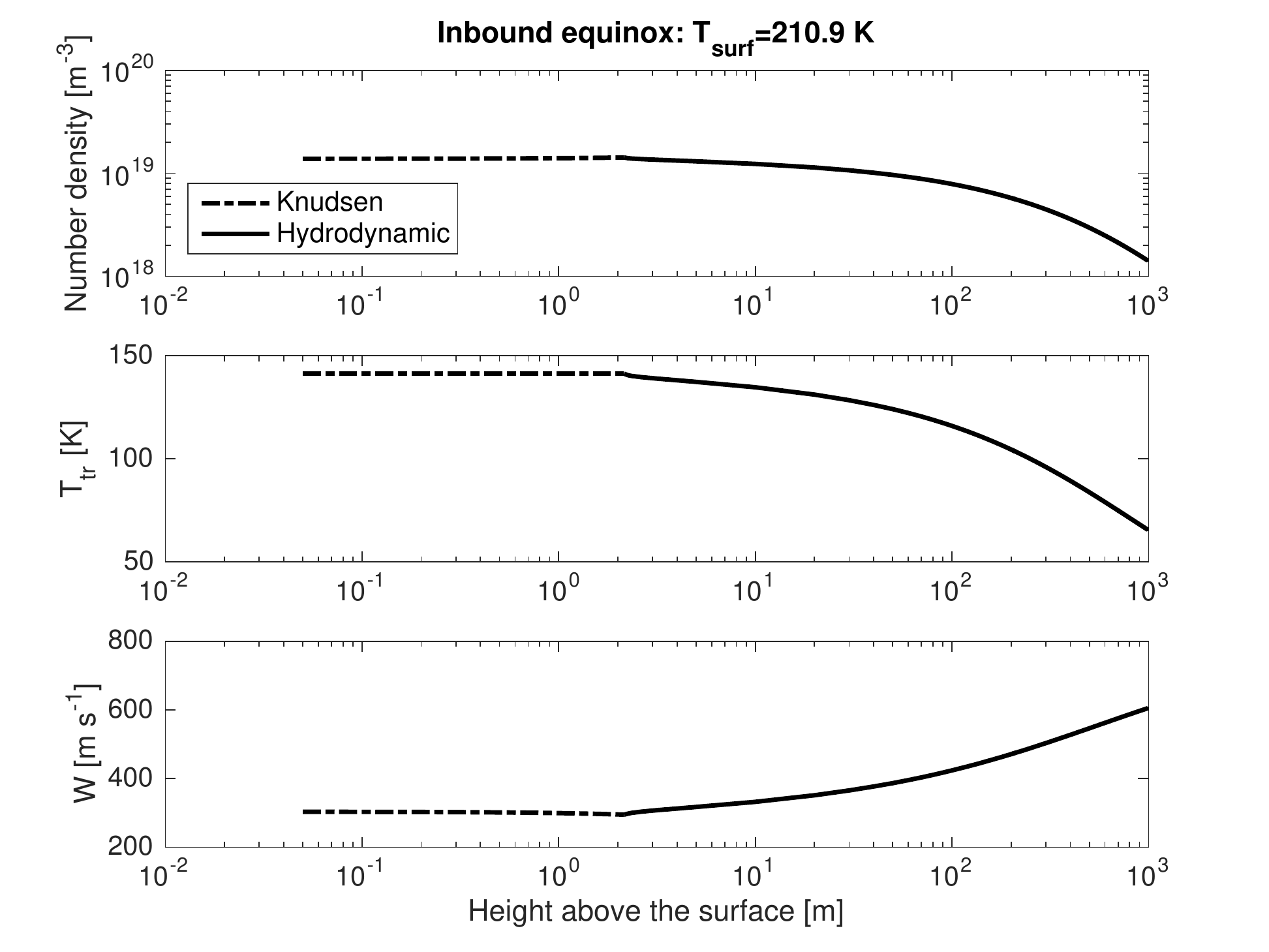}}
   \caption{Number density, translational temperature and drift speed at $T=211\,\mathrm{K}$.}
     \label{fig2}
\end{center}
\end{figure}

\noindent
Figure~\ref{fig2} shows the number density, translational temperature, and expansion velocity of the gas emanating from the 
same region as in Fig.~\ref{fig1} near noon (see the right circle in that figure). Due to the high gas production rate, this case 
illustrates the strong sublimation case described in Sec.~\ref{sec_method_coma_strong}, i.e., a Knudsen layer separates the nucleus 
and the hydrodynamic coma. Here, the Knudsen layer is $\sim 2.15\,\mathrm{m}$ thick. The initial gas drift speed is $\sim 303\,\mathrm{m\,s^{-1}}$ and 
the gas has an initial translational temperature of $\sim 141\,\mathrm{K}$, much lower than the surface temperature because of the 
kinetic jump condition (Eq.~\ref{eq:22}). As the gas recedes from the nucleus surface, the expansion results in a number density reduction 
which is nearly an order of magnitude at $1\,\mathrm{km}$ from the surface. The gas acceleration to $\sim 605\,\mathrm{m\,s^{-1}}$ is 
a consequence of mass conservation (a thinner gas needs to flow faster to carry the same amount of mass over a boundary). The 
adiabatic cooling of the gas is evident and the translational temperature drops to $\sim 66\,\mathrm{K}$ within the lower kilometer of the coma.\\

\begin{figure}
\begin{center}
     \scalebox{0.6}{\includegraphics{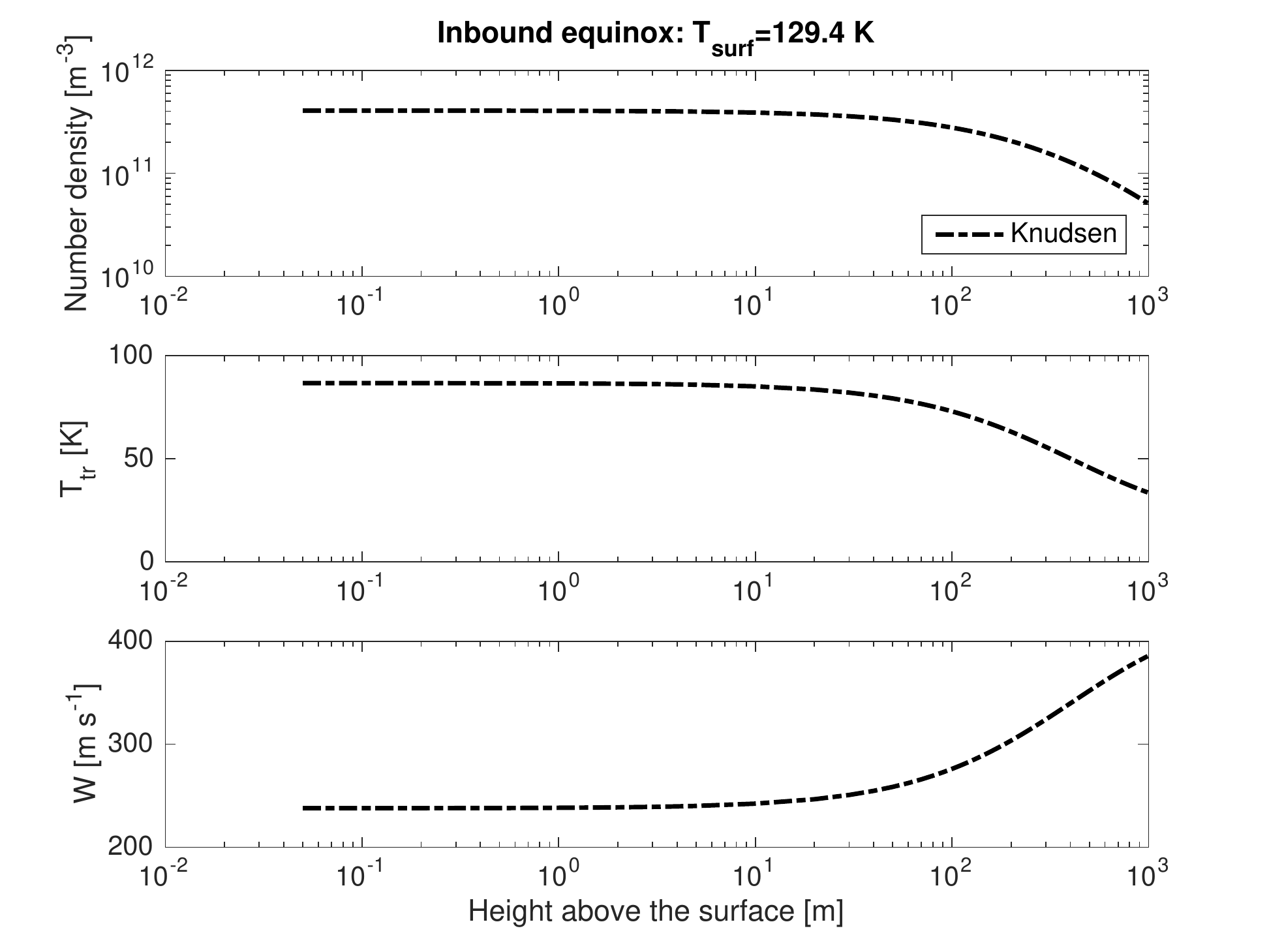}}
   \caption{Number density, translational temperature and drift speed at $T=129\,\mathrm{K}$.}
     \label{fig3}
\end{center}
\end{figure}

\noindent
The conditions in the nighttime coma are drastically different. Figure~\ref{fig3} shows the number density, translational temperature, and expansion velocity of the gas emanating from the 
same region as in Fig.~\ref{fig1} just before dawn (see the left circle in that figure). Because of the low production rate, Fig.~\ref{fig3} illustrates 
the weak sublimation case described in Sec.~\ref{sec_method_coma_weak}. The coma never achieves thermodynamic equilibrium and has to be modeled as an 
infinite Knudsen layer. Over the inner $1\,\mathrm{km}$ of the coma, the gas translational temperature drops from $\sim 87\,\mathrm{K}$ to  $\sim 34\,\mathrm{K}$ while 
the expansion velocity increases from $\sim 238\,\mathrm{m\,s^{-1}}$ to $\sim 385\,\mathrm{m\,s^{-1}}$. The number density is $\sim 7$ orders of magnitude lower than 
on the dayside, as expected.\\

\begin{figure}
\begin{center}
     \scalebox{0.6}{\includegraphics{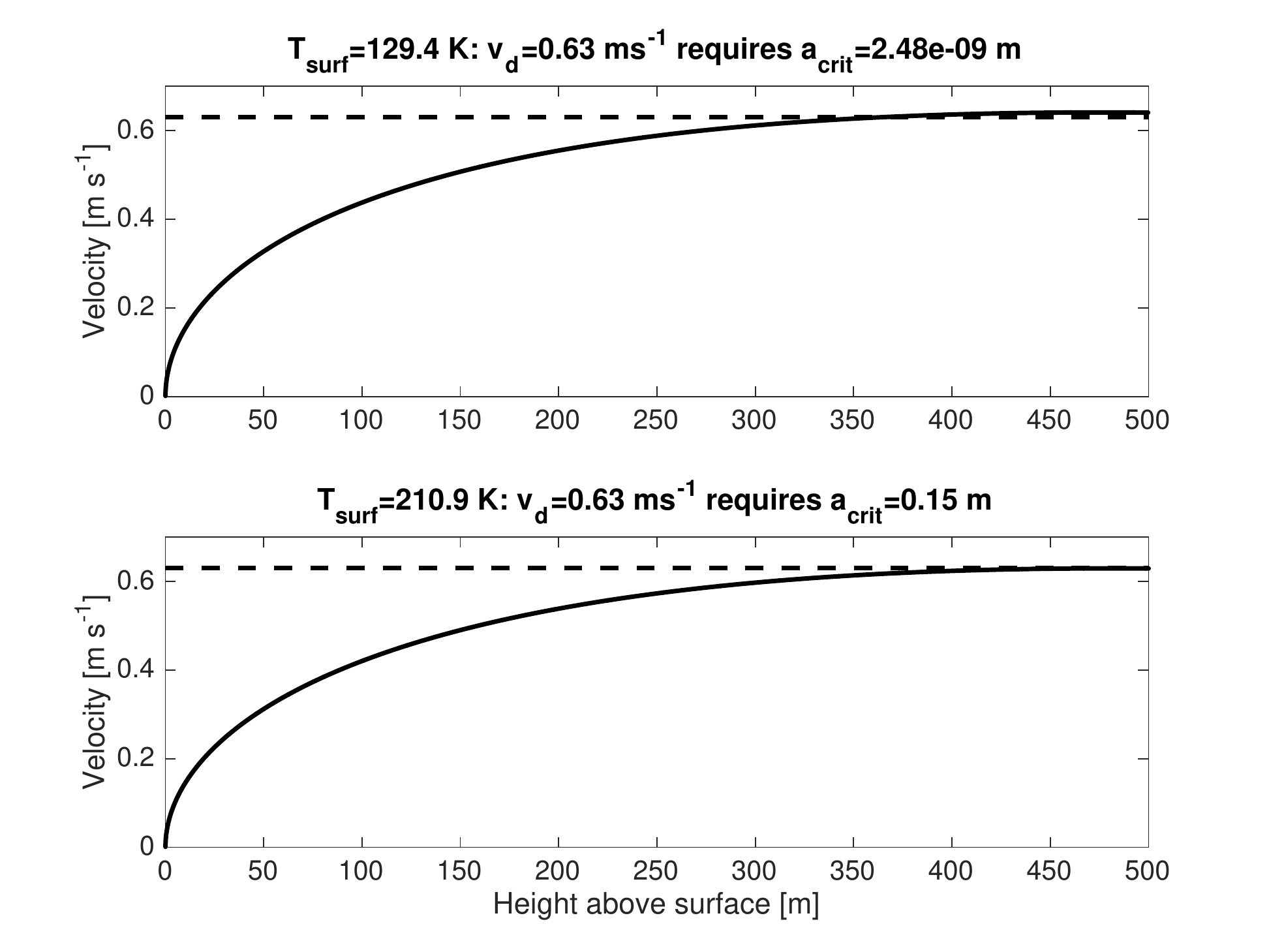}}
   \caption{Velocity versus height for chunks with about the right size to reach a targeted 
terminal velocity $v_{\rm d}$ (red) needed to take them to a given target region. The upper panel shows conditions near midnight while the 
lower panel shows conditions near noon.}
     \label{fig4}
\end{center}
\end{figure}

\noindent
The source region in question has such a position, local rotational velocity in the inertial frame, and surface normal that it would be capable of 
launching chunks to a particular target area on the opposite nucleus hemisphere if these chunks are accelerated to 
$v_{\rm d}\approx 0.63\,\mathrm{m\,s^{-1}}$ in the vicinity of the nucleus. For comparison, the escape velocity of 67P is $v_{\rm esc}=\sqrt{2G_{\rm N}M_{\rm n}/R_{\rm n}}=0.83\,\mathrm{m\,s^{-1}}$, 
where the nucleus mass is $M_{\rm n}=9.982\cdot 10^{12}\,\mathrm{kg}$ \shortcite{patzoldetal16} and the area--equivalent nucleus radius is $R_{\rm n}=1932\,\mathrm{m}$ \shortcite{jordaetal16}. The size of the chunks that will reach  $v_{\rm d}\approx 0.63\,\mathrm{m\,s^{-1}}$ (see Sec.~\ref{sec_method_eject}) 
varies drastically during the day. Figure~\ref{fig4} shows results from the numerical integration of Eq.~(\ref{eq:32}) for both the noon and near--dawn conditions discussed 
previously. At noon, the gas flow is sufficiently strong to lift chunks with diameters just under $1.9\,\mathrm{m}$. However, those chunks are moving too slowly to reach the 
particular target area in question. Chunks with $\sim 0.3\,\mathrm{m}$ diameter do reach the desired velocity (Fig.~\ref{fig4}, lower panel) and are expected to reach the target area, while 
chunks smaller than this will travel too fast. When the gas production rate wanes after noon, the critical chunk radius $a_{\rm crit}$ capable of entering the 
correct transport route to the target area diminishes rapidly. Figure~\ref{fig4} (upper panel) shows that only nano--sized grains would reach the desired velocity, if such particles exist.\\

\noindent
Throughout the orbit the capability of this and other source regions to launch large chunks towards the northern hemisphere changes significantly.  
Seen from many source regions, the Sun is circumpolar at perihelion and will continuously send large chunks to the north. For the region discussed above the solar 
radiation intensity actually diminishes because of its location on the nucleus. The surface temperature fluctuates between $147$--$206\,\mathrm{K}$ and 
gas release can eject chunks with up to $1.3\,\mathrm{m}$ diameter range. It would launch chunks with $v_{\rm d}\approx 0.63\,\mathrm{m\,s^{-1}}$ for 
diameters in the $1.4\,\mathrm{\mu m}$--$0.2\,\mathrm{m}$ range. These numbers serve the purpose of illustrating the drastic variations in airfall chunk sizes a 
given target region may experience in just a few months during perihelion approach, because of changing illumination conditions at the source during 
nucleus rotation and orbital motion.\\

\begin{figure}
\begin{center}
     \scalebox{0.6}{\includegraphics{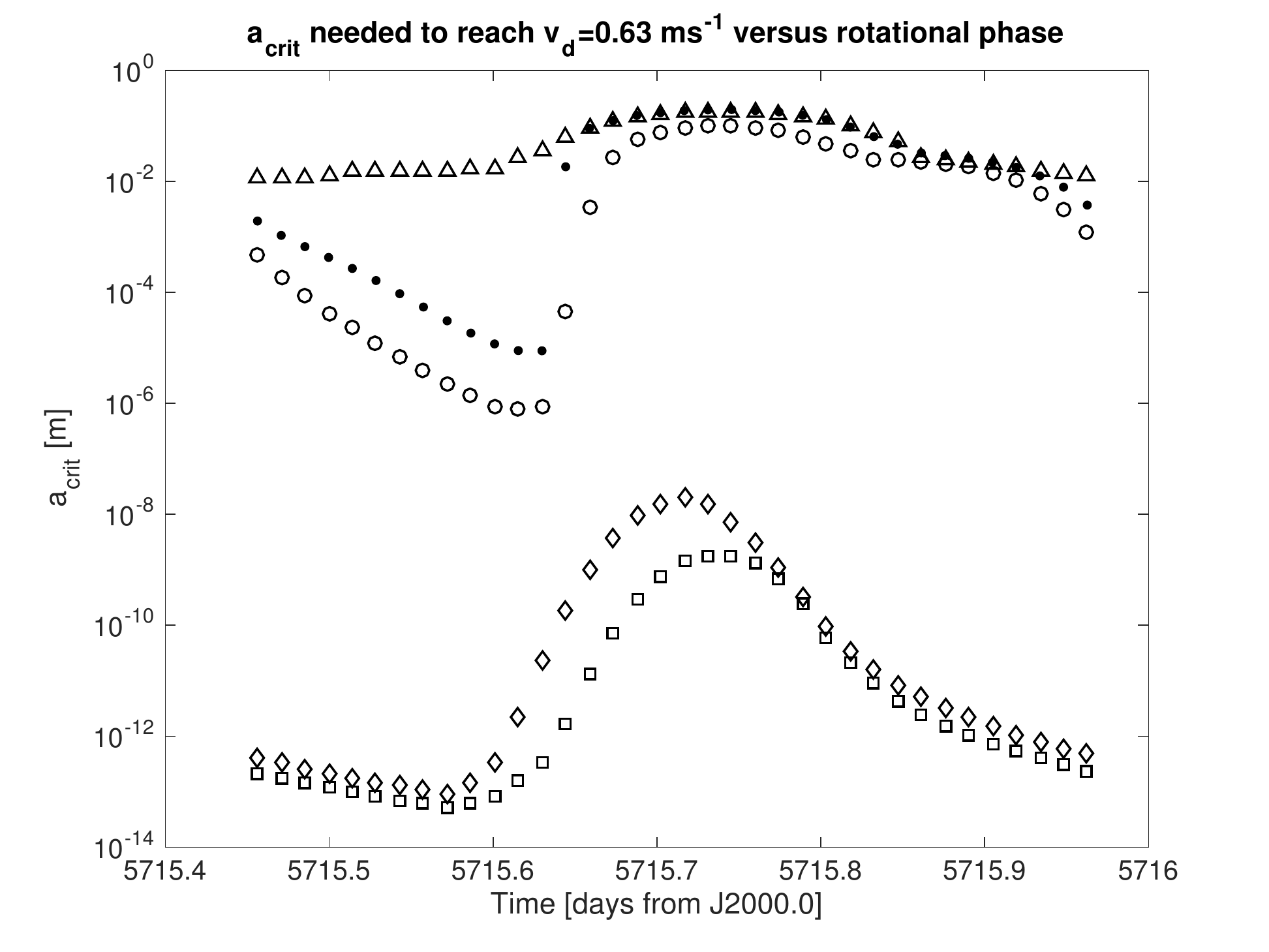}}
   \caption{Ejected chunk size versus time during one nucleus rotation near perihelion for a number of source 
regions (shown with different colors) that all feed the same target area.}
     \label{fig5}
\end{center}
\end{figure}

\noindent
The substantial variability of the critical chunk size during the comet day is shown more clearly in Fig.~\ref{fig5}. Here, $a_{\rm crit}(t)$ is shown for a 
number of different source regions that all contribute to target area \#3 (see Table~\ref{tab1}), for a single nucleus rotation near perihelion. One source is continuously 
feeding cm--to dm--sized chunks to the target area because the Sun is circumpolar at that location. A couple of other sources experience day/night 
variations and their contributions fluctuate between micron and $\sim 0.1\,\mathrm{mm}$ contributions at night to cm--to dm--sized chunks in the 
day. Other sources are poorly illuminated and are only capable of providing sub--micron particles. The material feeding a given target area therefore 
originates from a variety of locations and arrives in chunks of drastically different size.\\

\begin{figure}
\centering
\begin{tabular}{cc}
\scalebox{0.4}{\includegraphics{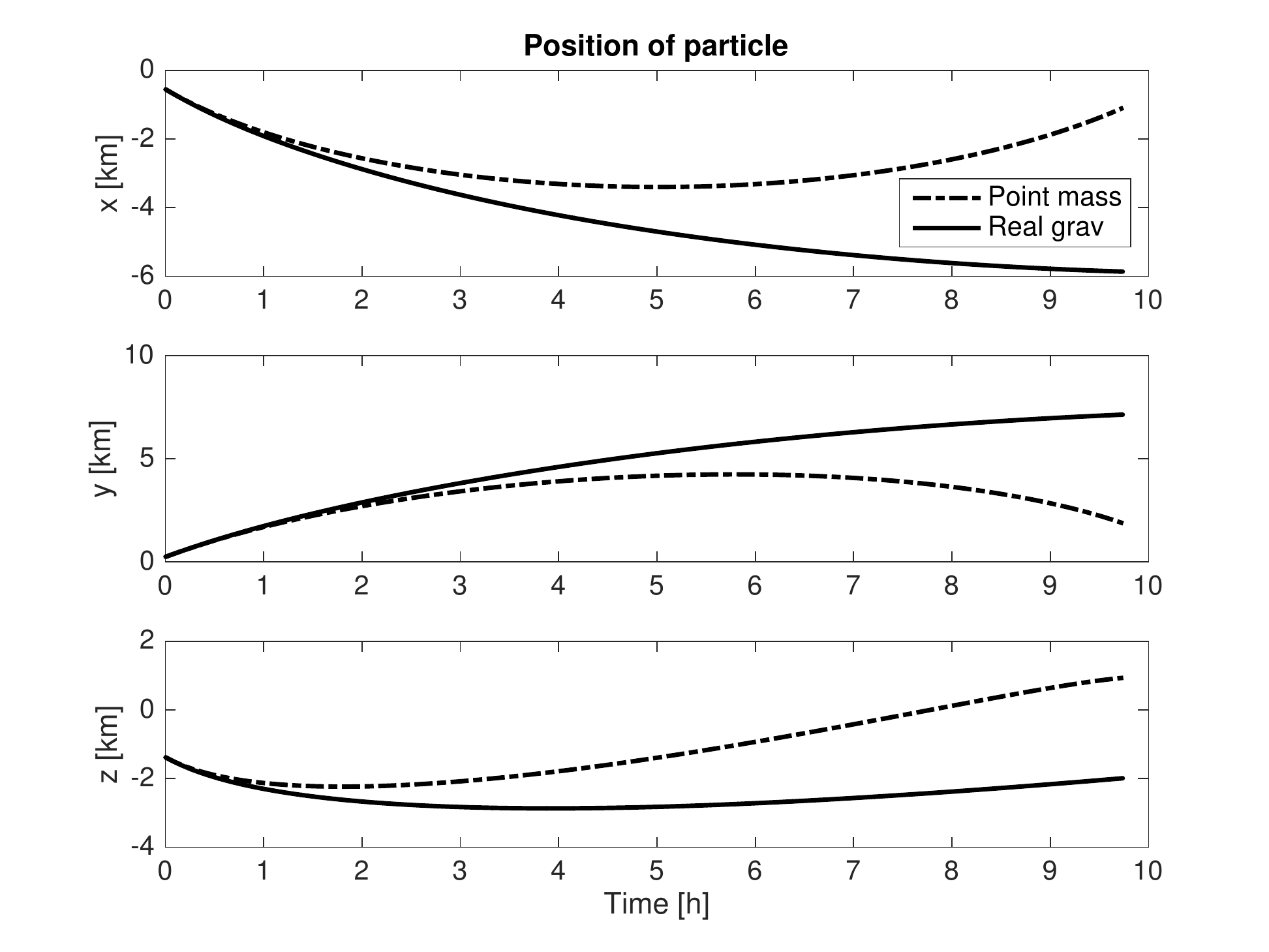}} & \scalebox{0.4}{\includegraphics{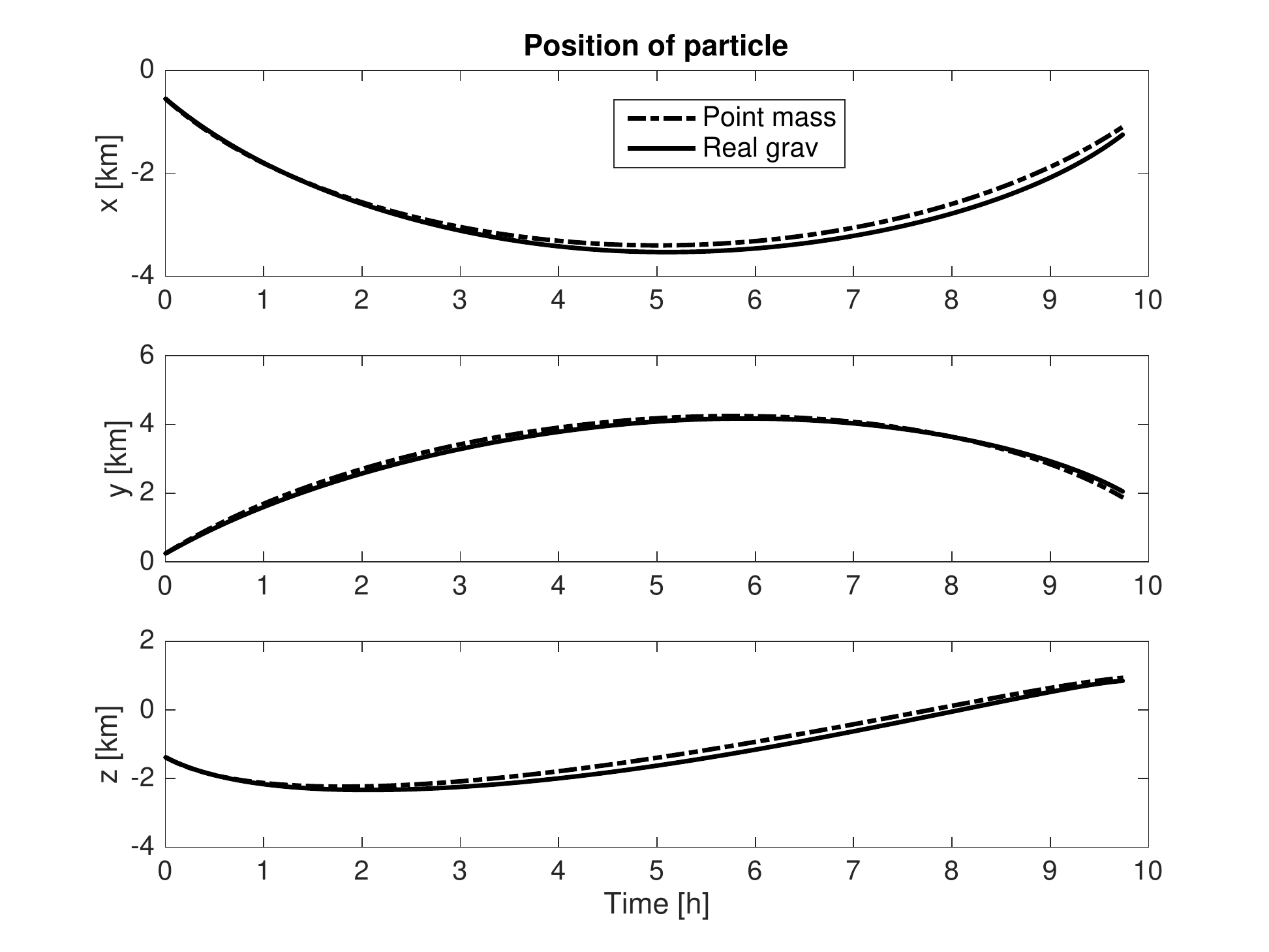}}\\
\end{tabular}
     \caption{Ecliptic Cartesian position coordinates versus time for a Keplerian trajectory connecting a source region with target area \#24 for an 
initial speed $v_{\rm d}=0.855\,\mathrm{m\,s^{-1}}$ along the local surface normal in addition a $0.081\,\mathrm{m\,s^{-1}}$ component due to 
nucleus rotation (black). The red curves correspond to a numerical orbit integration using a realistic gravity field of a rotating nucleus for $v_{\rm d}=0.855\,\mathrm{m\,s^{-1}}$ (left) 
and $v_{\rm d}=0.800\,\mathrm{m\,s^{-1}}$ (right).}
     \label{fig6}
\end{figure}

\noindent
Before proceeding to the main results of this paper we first discuss the accuracy of the orbital solutions for the chunks. The orbit calculations in this paper are made assuming a 
nucleus point mass and corresponding Keplerian orbits (Sec.~\ref{sec_method_kepler}). We now compare such solutions to numerically integrated orbits in the 
realistic gravity field of a rotating irregular nucleus (Sec.~\ref{sec_method_numorb}). The black curves in Fig.~\ref{fig6} shows the $\{x(t),\,y(t),\,z(t)\}$ coordinates in 
the ecliptic system of a Keplerian orbit that connected a particular source region with target area \#24. In this particular case, airfall at that location could be realized for 
$v_{\rm d}=0.855\,\mathrm{m\,s^{-1}}$ according to the idealized point mass solution and a time of flight of $9.7\,\mathrm{h}$. The left panel of Fig.~\ref{fig6} shows 
that the numerically integrated trajectory in a realistic gravity field rapidly diverges from the idealized one and after $9.7\,\mathrm{h}$ the difference amounts to $7.7\,\mathrm{km}$.\\

\noindent
The main reason for this drastic difference is that the nucleus potential at the launch site in question is merely $\sim 90\%$  of that obtained if all nucleus mass was 
concentrated at the origin. The chunk injects into an orbit that perhaps would have been similar to that around a point mass carrying $\sim 10\%$ less mass 
than 67P. If we consider the vis--viva equation (e.g., \shortciteNP{danby89}) that relates the orbital speed $v_{\rm o}$ and distance $r_{\rm o}$ from the origin for any 
elliptical orbit with semi--major axis $a_{\rm o}$,
\begin{equation} \label{eq:visviva}
v_{\rm o}^2=\mu_{\rm o}\left(\frac{2}{r_{\rm o}}-\frac{1}{a_{\rm o}}\right)
\end{equation}
we realize that any reduction of the reduced mass $\mu_{\rm o}$ to $\mu_{\rm o}^*<\mu_{\rm o}$ at a fixed $r_{\rm o}$ could be compensated for by lowering the velocity by a factor $\sqrt{\mu_{\rm o}^*/\mu_{\rm o}}$, thereby 
achieving exactly the same semi--major axis. Taking the velocity component due to nucleus rotation into account, this suggests that if the chunk velocity due to 
gas drag was reduced by $\sim 5\%$ compared to its original value (i.e., from $0.855\,\mathrm{m\,s^{-1}}$ to $0.800\,\mathrm{m\,s^{-1}}$) we would expect the numerically 
integrated chunk trajectory to follow the Keplerian one. The right panel of Fig.~\ref{fig6} shows a numerical orbit integration when $v_{\rm d}=0.8\,\mathrm{m\,s^{-1}}$ and 
the Keplerian orbit is indeed reproduced very well. It therefore seems like very modest modifications to the ejection speed could compensate for differences between the real 
and the idealized gravity field. In principle, an adjustment of the ejection velocity should be compensated for by a corresponding change of $a_{\rm crit}$. However, considering 
that the cross--section to mass ratio will vary among real chunks because they ought to have some density dispersion, we do not attempt such a correction but assume that 
our $a_{\rm crit}$--values are representative (particularly considering the factor 4 margin used in Sec.~\ref{sec_method_airfall} when evaluating the fraction of the ejected mass that 
contribute to airfall).\\

\begin{figure}
\begin{center}
     \scalebox{0.6}{\includegraphics{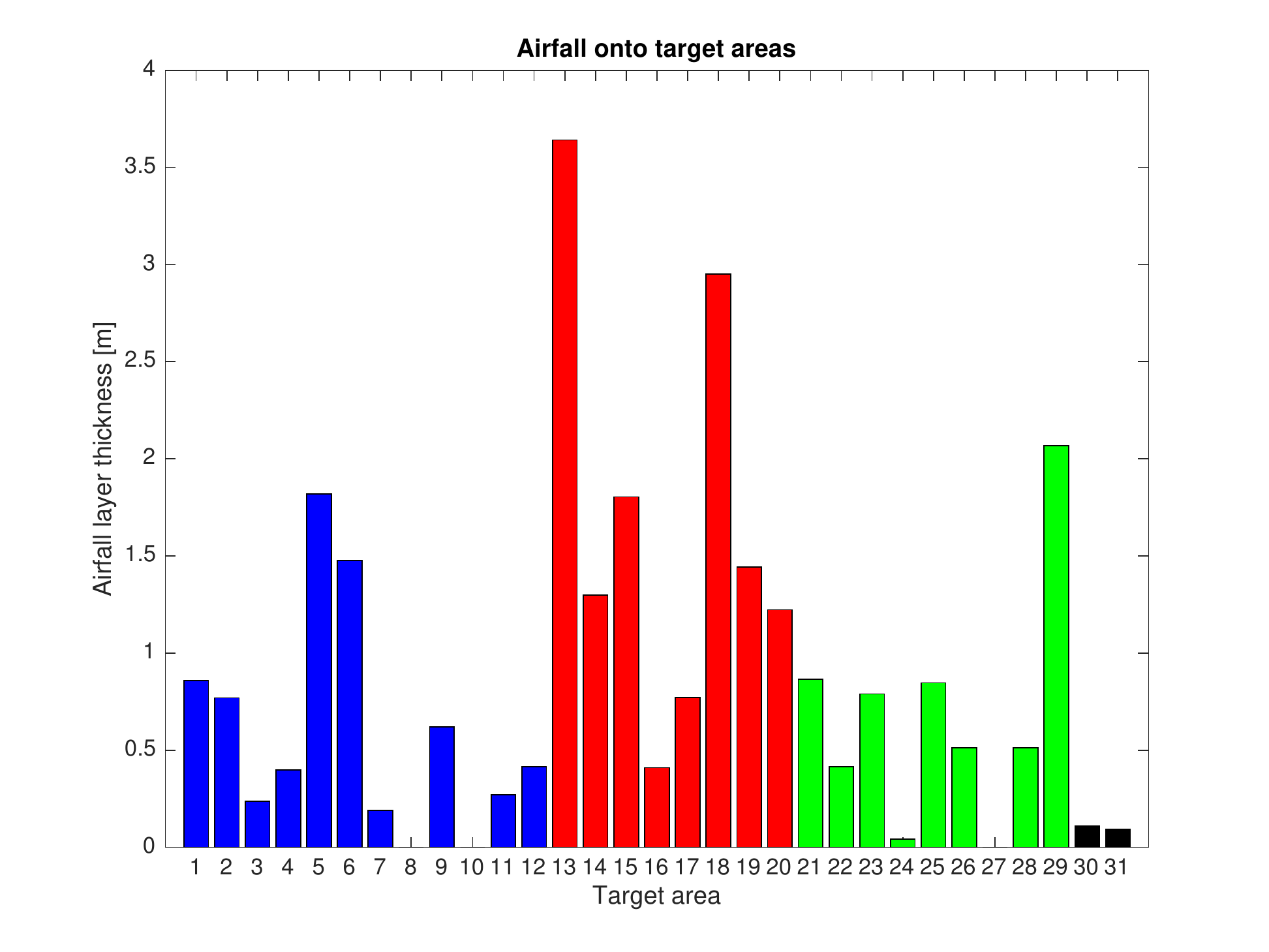}}
   \caption{The amount of accumulated airfall material expressed as layer thickness in meters for target areas in Serqet (blue), 
Ash (red and green), and Imhotep (black).}
     \label{fig7}
\end{center}
\end{figure}

\noindent
The final results of our simulations regarding airfall accumulation are tabulated in Table~\ref{tab1} and are shown graphically in Fig.~\ref{fig7}. The thicknesses of accumulated layers 
during a perihelion passage vary rather substantially across the nucleus surface. Some regions collect only small amounts of material, amounting to a few centimeters or less. Most 
regions accumulate a few decimeters. However, some regions (particularly \#5--6, 13--15, 18--20, and 29) collect substantial layers 
that are 1.5--$3.6\,\mathrm{m}$ thick. The average thickness for all 31 target areas is $0.87\,\mathrm{m}$. Target areas \#7, 8, and 10 are the ones located closest to Agilkia, 
where Philae bounced. The average airfall layer thickness for those target sites is $0.27\,\mathrm{m}$ according to our investigation. We therefore predict that the thickness of 
the airfall layer is comparably thin at this part of the nucleus (about a third of the average thickness) and we note that \shortciteN{bieleetal15} estimate a thickness of 
$\stackrel{>}{_{\sim}} 0.20\,\mathrm{m}$ for the ``granular soft surface'' that possibly is located ``on top of a more rigid layer''. We are encouraged by the similarity between 
the layer thickness inferred from the Philae measurements, and the estimate based on our calculations.\\

\noindent
Target areas \#1--12 are located in the Ma'at and Serqet regions on the small lobe, extending from the border of Hatmehit towards the Hathor cliff. Collectively, these areas 
accumulate an average of $0.6\,\mathrm{m}$, with the largest deposits (\#5--6) found at the Hatmehit border. Target areas \#13--20 are located in the eastern Ash region 
on the large lobe and extend from the border with Aten westward up to the antimeridian. As a group they have the largest average layer thickness of $1.7\,\mathrm{m}$ 
with the thinnest layers (\#16 and \#17) located at the very center. Target areas \#21--29 are also located in Ash but in its western part stretching from the antimeridian 
up to the borders to Seth, Anubis, and Aten. The average layer thickness is $0.7\,\mathrm{m}$ for this group with the largest accumulation (\#29) near those borders. 
Finally, target areas \#30-31 are located in Imhotep, just south of the equatorial plane within the vast smooth region that dominates this morphological unit. Because of 
efficient self--cleaning the net accumulation here is the lowest, $0.1\,\mathrm{m}$.\\

\begin{table}
\begin{center} {\bf } \end{center}
\begin{center}
\begin{tabular}{||l|r|r|r||l|r|r|r||}
\hline
\hline
Target & Lat & Long & Thickness $\mathrm{[m]}$ & Target &  Lat & Long & Thickness $\mathrm{[m]}$\\
\hline
\#1 & $40.9^{\circ}$ &  $338.2^{\circ}$ &  0.86 & \#17 & $49.6^{\circ}$ & $158.1^{\circ}$ &    0.77\\ 
\#2 & $34.6^{\circ}$ & $320.2^{\circ}$ & 0.77  &  \#18 & $58.1^{\circ}$ & $171.1^{\circ}$ &  2.95\\ 
\#3 & $30.3^{\circ}$ & $332.9^{\circ}$ &  0.24 & \#19 & $49.2^{\circ}$ & $177.1^{\circ}$ &  1.44\\ 
\#4 & $27.4^{\circ}$ & $340.2^{\circ}$ & 0.40 & \#20 & $42.5^{\circ}$ & $180.5^{\circ}$ & 1.22\\ 
\#5 & $19.8^{\circ}$ & $346.0^{\circ}$ & 1.82 & \#21 & $33.9^{\circ}$ & $203.1^{\circ}$ &  0.87\\  
\#6 & $9.1^{\circ}$ & $349.2^{\circ}$ &  1.48 &  \#22 & $30.3^{\circ}$ & $200.9^{\circ}$ & 0.42\\
\#7 & $19.0^{\circ}$ & $339.9^{\circ}$ &  0.19 & \#23 & $25.4^{\circ}$ & $206.4^{\circ}$ &  0.79\\
\#8 & $16.4^{\circ}$ & $336.0^{\circ}$ & 0.00 &  \#24 & $24.4^{\circ}$ & $198.8^{\circ}$ & 0.04\\
\#9 & $21.2^{\circ}$ & $331.7^{\circ}$ &  0.62 &   \#25 & $19.1^{\circ}$ & $195.9^{\circ}$ &  0.85\\
\#10 & $13.9^{\circ}$ & $330.6^{\circ}$ & 0.00 & \#26 & $17.6^{\circ}$ & $202.9^{\circ}$ & 0.51\\
\#11 & $6.4^{\circ}$ & $333.5^{\circ}$ &  0.27 &  \#27 & $19.2^{\circ}$ & $206.1^{\circ}$ & 0.00\\
\#12 & $15.3^{\circ}$ & $315.9^{\circ}$ &  0.41 &   \#28 & $16.8^{\circ}$ & $212.1^{\circ}$ &  0.51\\
\#13 & $48.87^{\circ}$ &  $108.0^{\circ}$ &  3.64 &  \#29 & $11.3^{\circ}$ & $210.1^{\circ}$ &  2.07\\
\#14 & $56.5^{\circ}$ & $120.9^{\circ}$ &  1.30 &  \#30 & $-10.7^{\circ}$ & $118.6^{\circ}$ &  0.11\\
 \#15 & $45.7^{\circ}$ & $135.6^{\circ}$ &  1.80 &  \#31 & $-7.2^{\circ}$ & $146.1^{\circ}$ &  0.09\\
\#16 & $58.9^{\circ}$ & $142.4^{\circ}$ & 0.41 &  & & & \\
\hline 
\hline
\end{tabular}
\caption{Target area locations and thickness of accumulated layers formed by fallback of material released from the southern hemisphere.}
\label{tab1}
\end{center}
\end{table}

\subsection{Volatile loss during the transfer through the coma} \label{sec_results_iceloss}

We now proceed to the question of the ice abundance in airfall deposits. Observations of dust jet switch--off after local sunset shows that the 
water ice sublimation front is located just a few millimeters below the nucleus surface \shortcite{shietal16}. The process that liberates and ejects 
chunks in the cm--m class, including but not limited to various forms of cracking (\shortciteNP{elmaarryetal15b}; \shortciteNP{augeretal18}) and cliff collapse 
\shortcite{pajolaetal17}, combined with gas drag, should produce coma chunks with an initial ice abundance that may be similar to that of the bulk nucleus. 
However, some of this ice will be lost during the flight through the coma, thereby lessening the initial ice abundance in airfall deposits.\\

\noindent
In order to estimate the ice abundance in fresh airfall deposits, we used the code \texttt{NIMBUS} described in Sec.~\ref{sec_method_grainthermo} 
to study a $D=0.1\,\mathrm{m}$ chunk in the coma, rotating about a fixed axis with coordinates $\{\alpha,\,\delta\}=\{0^{\circ},\,45^{\circ}\}$ in the equatorial 
system with a period $P=20\,\mathrm{min}$ (individual chunks with similar sizes and rotation periods have been observed in the 67P coma, see \shortciteNP{davidssonetal15b}). 
These simulations assumed a porosity of $70\%$, micro--meter sized grains (influencing the volume sublimation rate $q$) and pore lengths and radii of $10\,\mathrm{\mu m}$ 
and $1\,\mathrm{\mu m}$, respectively (influencing the gas diffusivities $\Phi$ and $\Psi$ as well as the magnitude of radiative heat transfer). The chunk was resolved by 18 latitudinal segments and 69 cells in the radial direction 
using geometric progression such that surface cells were $0.5\,\mathrm{mm}$ thick 
and core cells were $1\,\mathrm{mm}$ thick. The diurnal skin depth was resolved by $\sim 10$ cells.\\

\noindent
After a $12\,\mathrm{h}$ flight this body lost a total of $4.8\%$ of its 
original ice content. Because parts of the chunk received little radiation due to its spin axis orientation, and because small bodies of this type may not maintain a 
fixed spin axis, we then introduced a constant precession rate of $\alpha$ from $0^{\circ}$ to $360^{\circ}$ in one hour, combined with a declination nutation with a 
full oscillation from $\delta=90^{\circ}$ to $-90^{\circ}$ and back again during the same amount of time. Because of the more even distribution of illumination, the volatile 
loss increased marginally to $6.4\%$ of the initial ice content.\\

\noindent
Next, a $D=0.01\,\mathrm{m}$ chunk was considered. The rotation period was reduced to $2\,\mathrm{min}$ but the same precession and nutation periods were applied as before. 
Because of the smaller size of this chunk its ice loss was more substantial. Yet after $3\,\mathrm{h}$, only $37\%$ of the ice had been lost, increasing to $45\%$ loss after $6\,\mathrm{h}$ and to 
$56\%$ loss after $12\,\mathrm{h}$. This illustrates that the ice--loss rate slows down rapidly once the sublimation front has moved a few millimeters below the surface, and 
that chunks as small as $\sim 1\,\mathrm{cm}$ are capable of retaining a substantial fraction of their water ice on time--scales similar to the coma flight time. Coupled with the 
observation that the majority of mass in smooth region deposits is bound in $D\stackrel{>}{_{\sim}} 1\,\mathrm{cm}$ chunks (with at least half of that mass in 
$D\stackrel{>}{_{\sim}} 0.1\,\mathrm{m}$ chunks) this means that fresh northern airfall deposits are nearly as icy as the southern source regions that produced them.\\

\begin{figure}
\centering
\begin{tabular}{cc}
\scalebox{0.4}{\includegraphics{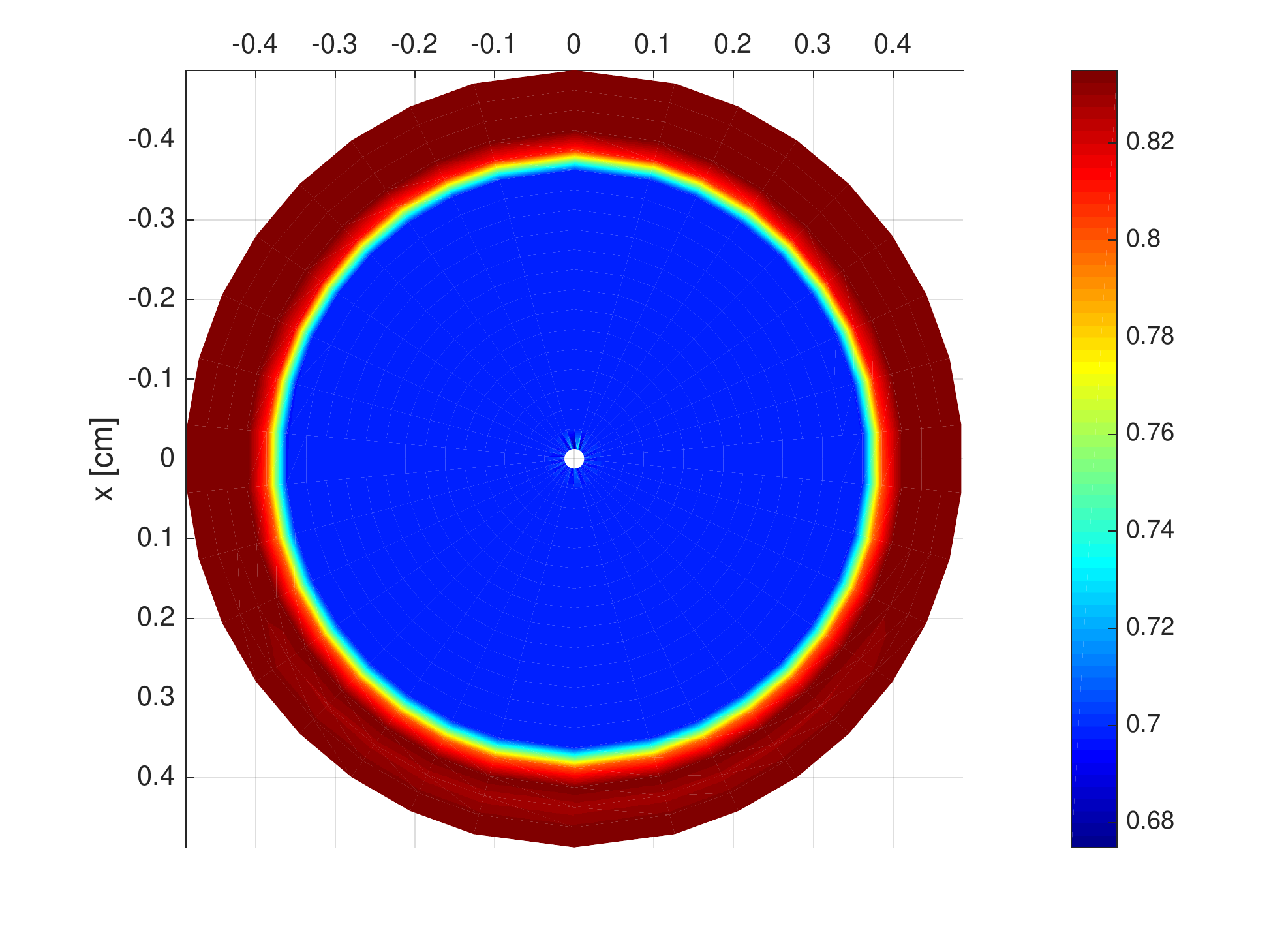}} & \scalebox{0.4}{\includegraphics{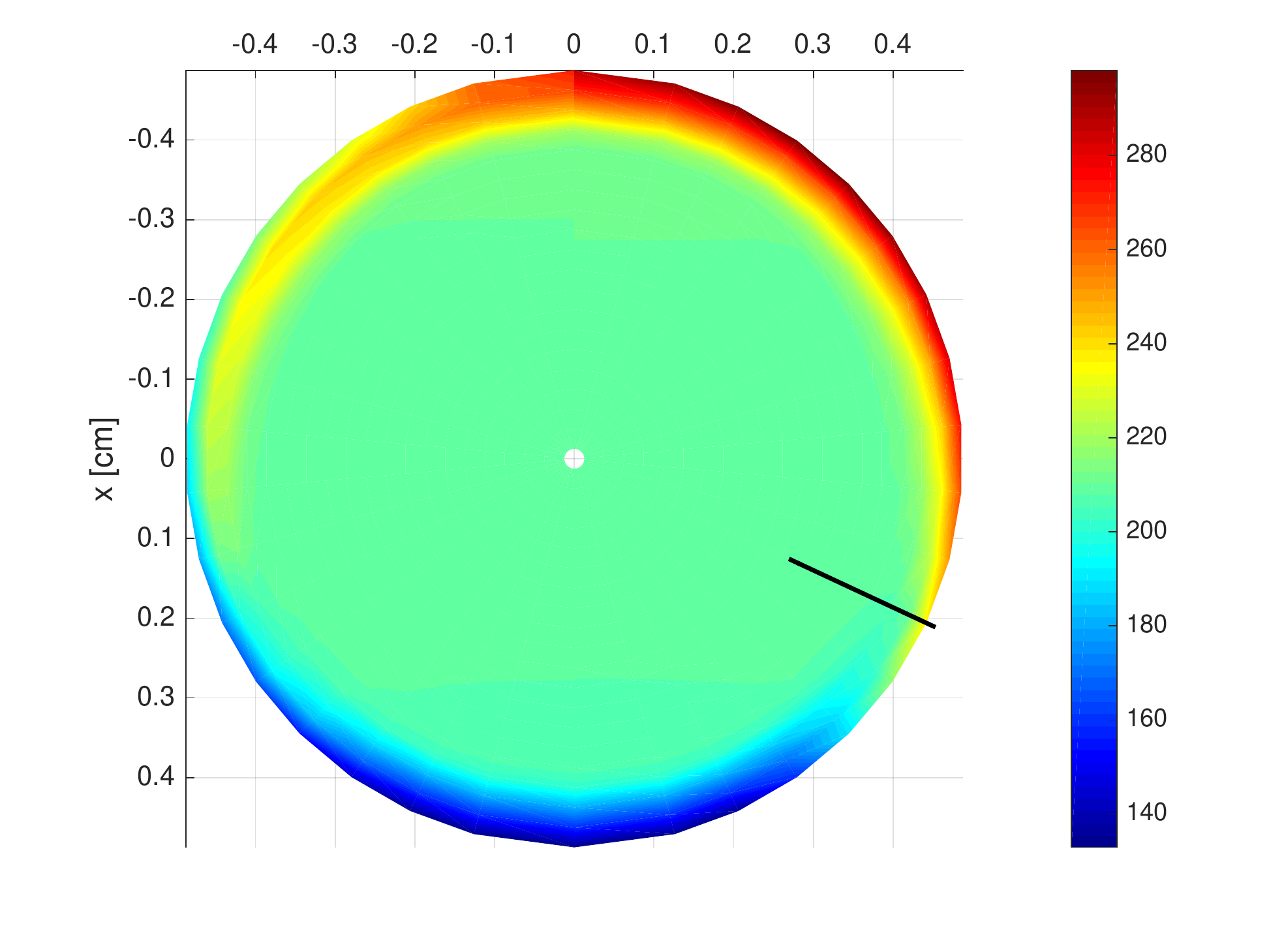}}\\
\scalebox{0.4}{\includegraphics{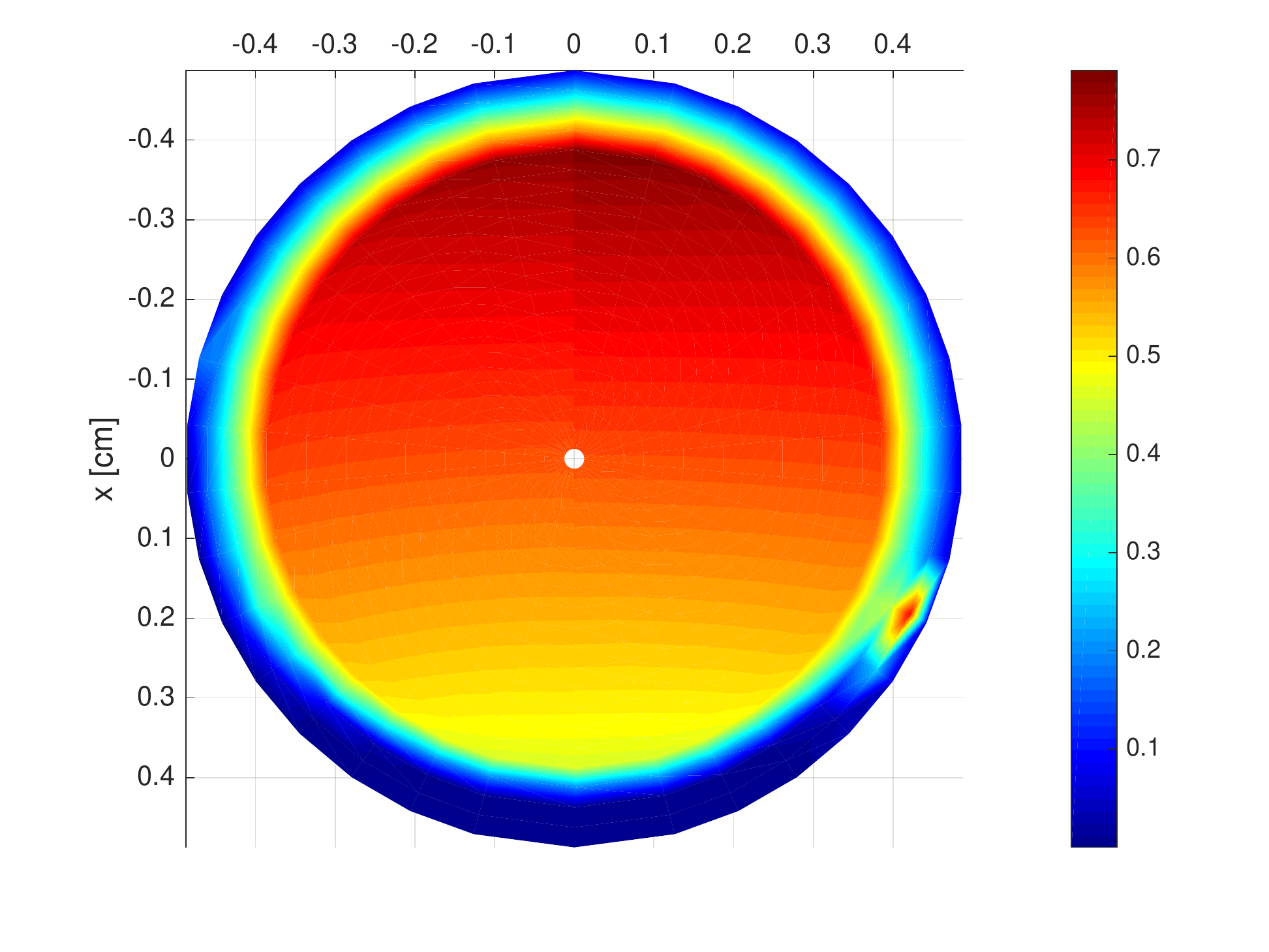}} & \scalebox{0.4}{\includegraphics{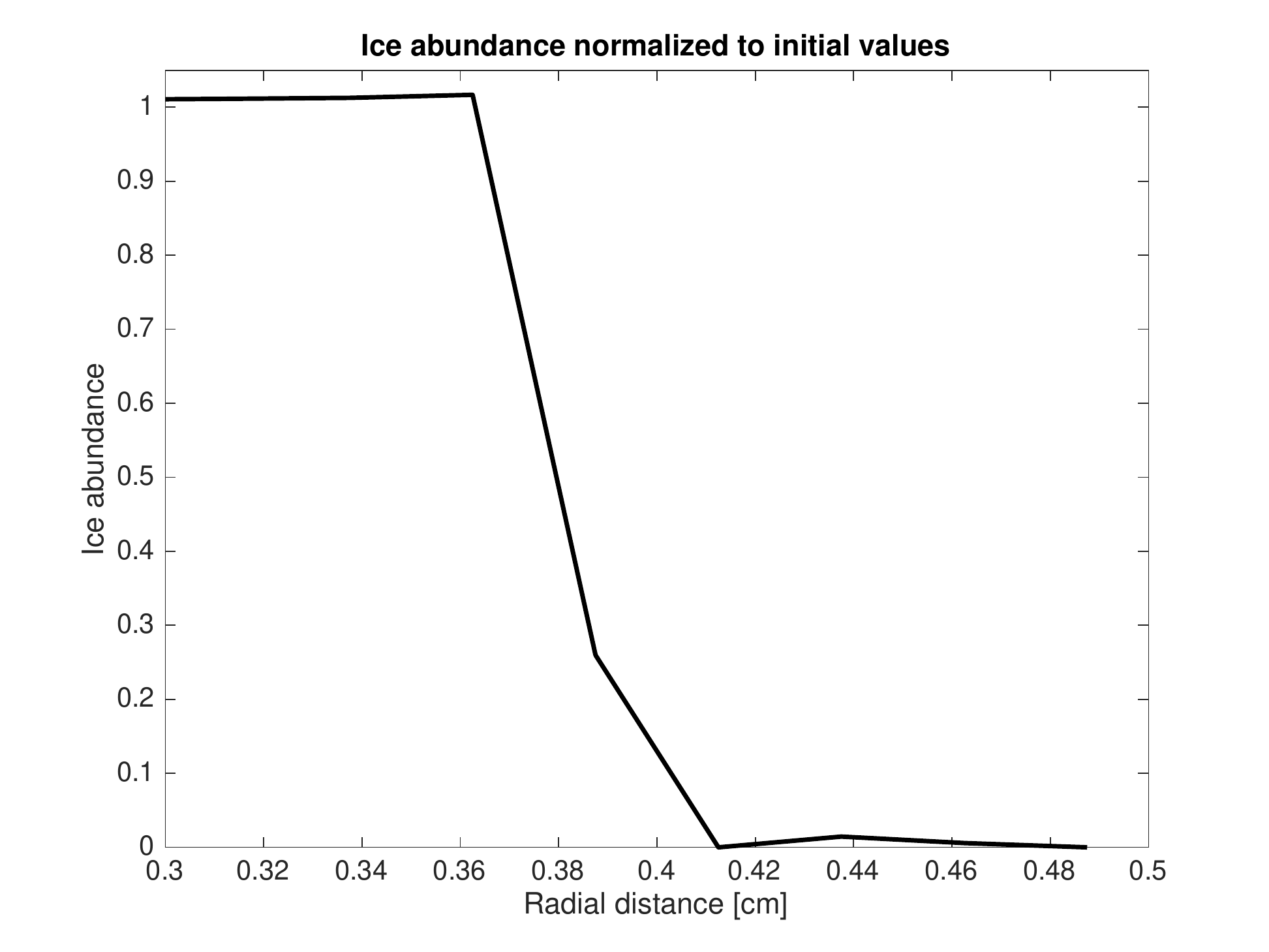}}\\
\end{tabular}
     \caption{Properties of a $D=1\,\mathrm{cm}$ coma chunk after $11.4\,\mathrm{h}$ in the coma. Upper left: porosity. Upper right: temperature $\mathrm{[K]}$. 
Lower left: $\mathrm{H_2O}$ vapor pressure $\mathrm{[Pa]}$. Lower right: ice abundance versus distance from the center at chunk latitude $25^{\circ}\,\mathrm{S}$, 
normalized to the initial local value. The location of the displayed region is shown as a solid black line in the upper right panel.}
     \label{fig8}
\end{figure}

\noindent
Figure~\ref{fig8} shows the internal distributions of porosity, temperature, and $\mathrm{H_2O}$ gas pressure for a $D=1\,\mathrm{cm}$ chunk about $11.4\,\mathrm{h}$ 
into the simulations. The sublimation front has withdrawn to a depth of $\sim 1\,\mathrm{mm}$ and the depth does not vary significantly with latitude because of the 
tumbling of the chunk that exposes all parts of the surface to strong solar illumination over time. At the particular snapshot shown in Fig.~\ref{fig8} the subsolar point 
has been at high northern latitudes sufficiently long for the south pole to cool off, and $\mathrm{H_2O}$ vapor has recondensed within the dust mantle. This is 
hinted at by a slight increase of porosity in the middle of the southern hemisphere dust mantle, and is seen more clearly in the lower right panel showing the ice 
abundance (normalized to the initial amounts) versus radial distance at latitude $25^{\circ}\,\mathrm{S}$. At depth the normalized ice abundance is slightly above unity 
because of downward diffusion and recondensation of vapor. Near the surface the ice abundance has previously gone to zero in the dust mantle, and recently been slightly 
elevated again, because of the previously mentioned near--surface cooling and associated vapor recondensation. The subsolar point is moving south which means 
that this frost is starting to sublimate at latitudes near the equator. That gives rise to the local vapor pressure maximum seen in the lower left panel.\\

\noindent
In order to explore the robustness of our results, we made three tests where the nominal parameters in the $D=0.1\,\mathrm{m}$ model were changed. 
First, the radius of constituent grains was changed from $r_{\rm g}=1\,\mathrm{\mu m}$ to $100\,\mathrm{\mu m}$, which decreases the volume sublimation rate by two orders 
of magnitude ($q\propto r_{\rm g}^{-1}$). The ice loss as well as internal temperatures and vapor pressures changed insignificantly. This can be understood 
as follows. During one second the part of the difference between absorbed flux and thermally emitted flux that is not used to heat sub--surface material must 
necessarily be consumed by net sublimation of ice. If an amount of energy $\Delta E$ is available, a mass $\Delta m=\Delta E/\mathcal{L}(T)$ must be converted to 
vapor (where the latent heat is evaluated at the temperature of the sublimation front). The temperature and pressure gradients on both sides of the 
sublimation front will evolve in such a way that gas diffusion removes $\Delta m$ from the sublimation front region. This flow gives rise to continuous 
reductions of the vapor pressure $p_{\rm v}$ below the local saturation pressure $p_{\rm sat}$, which triggers sublimation because $q\propto p_{\rm sat}(T)-p_{\rm v}$. 
The time--scale for re--establishing saturation conditions after a temporary deviation is extremely short (micro--seconds or less) because $q\gg0$ when $p_{\rm v}<p_{\rm sat}$ (and 
$q$ rapidly goes to zero when $p_{\rm v}\rightarrow p_{\rm sat}$). Therefore, even if the response time to pressure changes is increased two orders of magnitude 
because of $r_{\rm g}$, this time is still much shorter than the characteristic time scales of gas diffusion and heat conduction, thereby making the model insensitive to 
the size of the icy particles.\\

\noindent
In another test the dimensions of the cylindrical tubes used to evaluate gas diffusivity were increased by two orders of magnitude from radius $1\,\mathrm{\mu m}$ to $0.1\,\mathrm{mm}$ 
and from length $10\,\mathrm{\mu m}$ to $1\,\mathrm{mm}$. Note that the pore radius also affects the radiative component of heat conduction, but that parameter 
was still kept at $1\,\mathrm{\mu m}$ in order to isolate the dependence of the ice loss on the gas diffusion modeling. In this case, the water ice loss increased from $6.4\%$ 
to $8.9\%$. When the gas diffusivity increases drastically because of the longer and wider pores, the main effect is a reduction of temperature and vapor pressure, 
in such a way that the resulting gas fluxes inward and outward are essentially the same as before. The same net sublimation $\Delta m$ is required in order to 
consume the discrepancy between absorbed and emitted radiation that does not change much, and that goal can be met for a cooler and less pressurized gas. For example, a 
cell that had $T=204.4\,\mathrm{K}$ with the stricter tubes cooled by $15.4\,\mathrm{K}$ to $T=189.0\,\mathrm{K}$ when the tubes widened, and the vapor 
pressure fell from $p_{\rm v}=0.2234\,\mathrm{Pa}$ to $p_{\rm v}=0.0055\,\mathrm{Pa}$. The sub--surface cooling slightly reduces the surface temperatures as well, meaning that 
less energy is lost radiatively to space and more energy is available for sublimation. That is why the ice loss becomes somewhat larger, but this increase is still modest.\\

\noindent
In the final test, the pore radii used to regulate the radiative contribution to heat conduction was increased by two orders of magnitude as well. 
This is expected to have a stronger effect on the ice loss because the dust mantle becomes rather warm ($325\,\mathrm{K}$) compared to the 
icy interior, and wider pores allow for the thermal radiation to penetrate more efficiently to the ice. In fact, the total ice loss increased from 
$6.4\%$ to $13.7\%$. Yet, this increase is not large enough to alter our overall conclusions -- coma chunks preserve water ice rather efficiently.\\

\noindent
We do not expect these numbers to be strongly dependent on the dust--to--ice mass ratio. The water ice sublimation front is expected to 
withdraw to similar depths regardless of water abundance. At that critical depth, the cooling by net sublimation is reduced by the dust mantle quenching of gas diffusion 
to the point that the hot dust mantle dissipates the majority of absorbed energy through thermal reradiation. The fractional volatile losses in terms of volume and mass 
are thus expected to be similar.\\

\noindent
We now consider the presence and survival of $\mathrm{CO_2}$ ice in coma chunks. \shortciteN{davidssonetal20} reproduced the observed pre--perihelion $\mathrm{CO_2}$ 
production rate curve of 67P from $3.5\,\mathrm{AU}$ to $1.24\,\mathrm{AU}$ with a \texttt{NIMBUS} model that had the perihelion $\mathrm{CO_2}$ sublimation front located 
$\sim 1.9\,\mathrm{m}$ below the surface near the equator, and where that depth diminished with latitude to $\sim 0.2\,\mathrm{m}$ at the south pole. If the suggested presence of 
$\mathrm{CO_2}$ ice at shallow depths is correct, it is conceivable that regular sublimation--driven activity occasionally can  
expel extremely cold chunks containing supervolatiles at perihelion. The presence of $\mathrm{CO_2}$ ice patches on the surface of 67P observed 
spectroscopically by \shortciteN{filacchioneetal16} could be the result of such temporary exposure of near--surface $\mathrm{CO_2}$ ice. There are other forms of 
activity that might produce $\mathrm{CO_2}$--rich chunks as well. For example, the cliff collapse in Aswan observed by OSIRIS suddenly exposed material 
located $\sim 12\,\mathrm{m}$ below the previous surface \shortcite{pajolaetal17}, and though such events are rare, they could occasionally inject 
unusually cold chunks into the coma that still contain $\mathrm{CO_2}$. It is interesting and important to understand to what extent such material could be 
transported and mixed in with other airfall material.\\ 

\noindent
Therefore, in our last numerical experiment, 5\% condensed $\mathrm{CO_2}$ relative to $\mathrm{H_2O}$ (by number) was added to the chunk, and the initial 
temperature was lowered from $T(t=0)=150\,\mathrm{K}$ to $T(t=0)=50\,\mathrm{K}$ to avoid an explosive sublimation of the carbon dioxide. Our \texttt{NIMBUS} 
simulations of a $D=0.1\,\mathrm{m}$ chunk shows that it takes $2\,\mathrm{h}$ for all $\mathrm{CO_2}$ to be lost. A substantially larger chunk could potentially 
preserve a fraction of its $\mathrm{CO_2}$ ice during the transfer to the northern hemisphere.  After a total of $12\,\mathrm{h}$ exposure to 
the Sun at the perihelion distance of 67P, the amount of water ice lost in the $D=0.1\,\mathrm{m}$ chunk is $5.4\%$, i.e., somewhat lower than in the simulation without $\mathrm{CO_2}$. 
This reduction happens for two reasons; 1) more energy is required to heat the body which start out $100\,\mathrm{K}$ colder than in the other simulations; 
2) energy is needed in order to sublimate the carbon dioxide. As a consequence, the water sublimation is somewhat delayed.

\subsection{Ice loss from airfall deposits} \label{sec_results_survivability}

To perform the illumination calculations (including shadowing and self heating effects from the irregular nucleus) and the 
thermophysical simulations for the entire orbit, with rotation resolved (solving the thermophysical equations, Eq.~\ref{eq:06}--\ref{eq:09} 
requires a time step of $\sim 10\,\mathrm{s}$) is computationally demanding. Therefore, we have not performed these calculations 
for all 31 target areas, but for a subset of four. These areas were selected to represent each of the major regions under study. 
Specifically, we consider \#3 in Ma'at on the small lobe, \#16 in eastern Ash, \#24 in western Ash, and \#31 in Imhotep on the large lobe.\\

\begin{figure}
\centering
\begin{tabular}{cc}
\scalebox{0.4}{\includegraphics{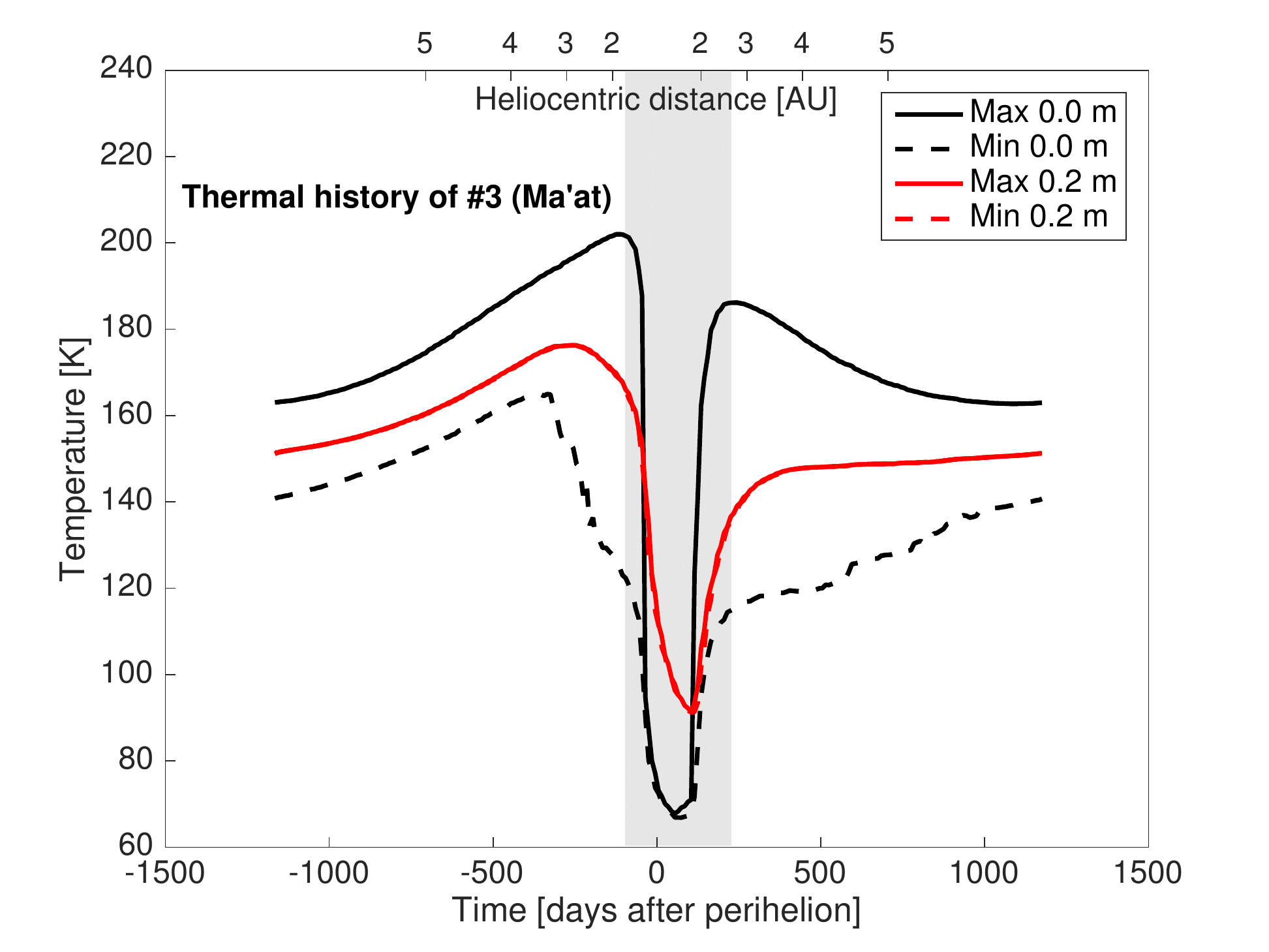}} & \scalebox{0.4}{\includegraphics{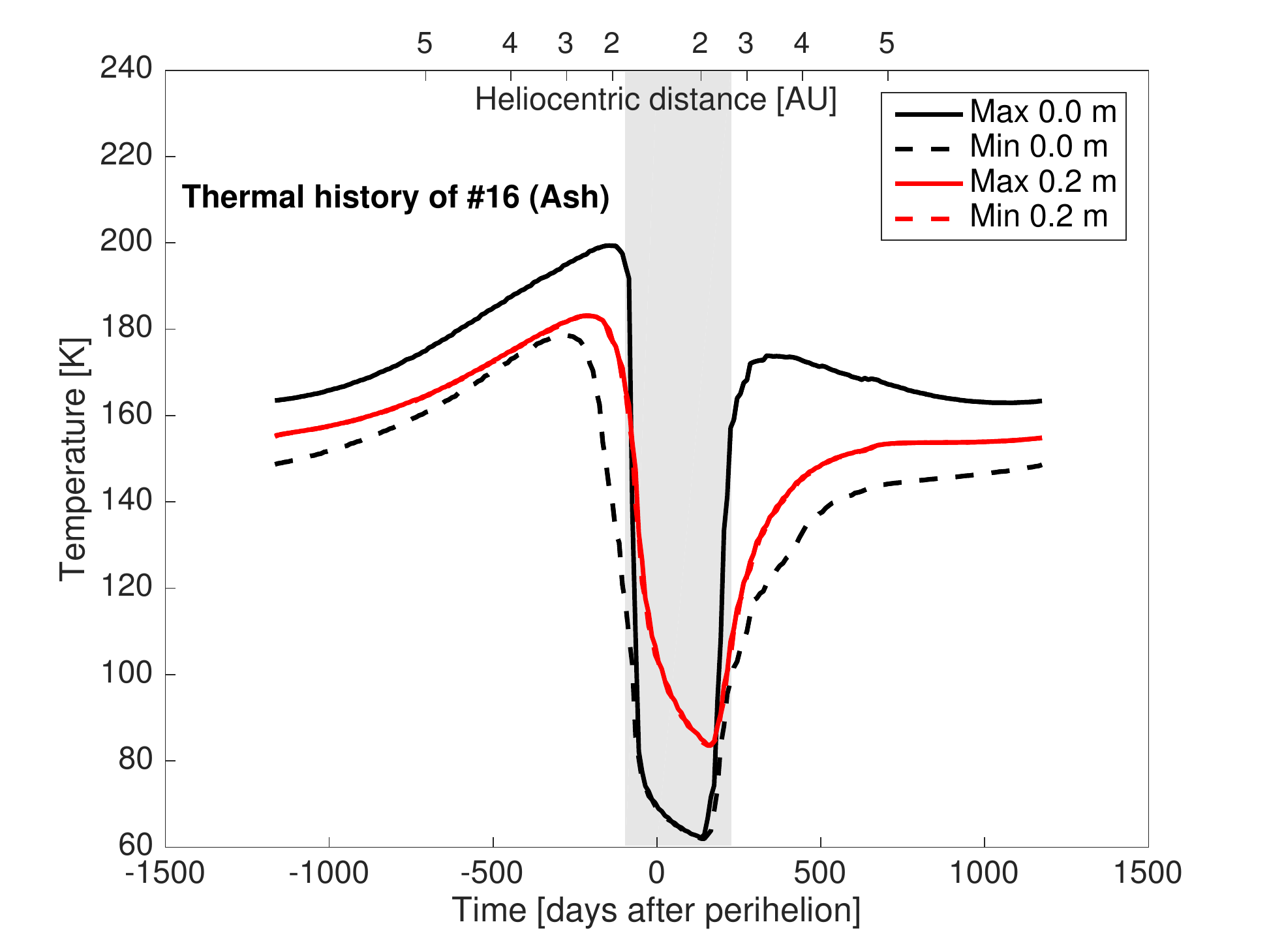}}\\

\scalebox{0.4}{\includegraphics{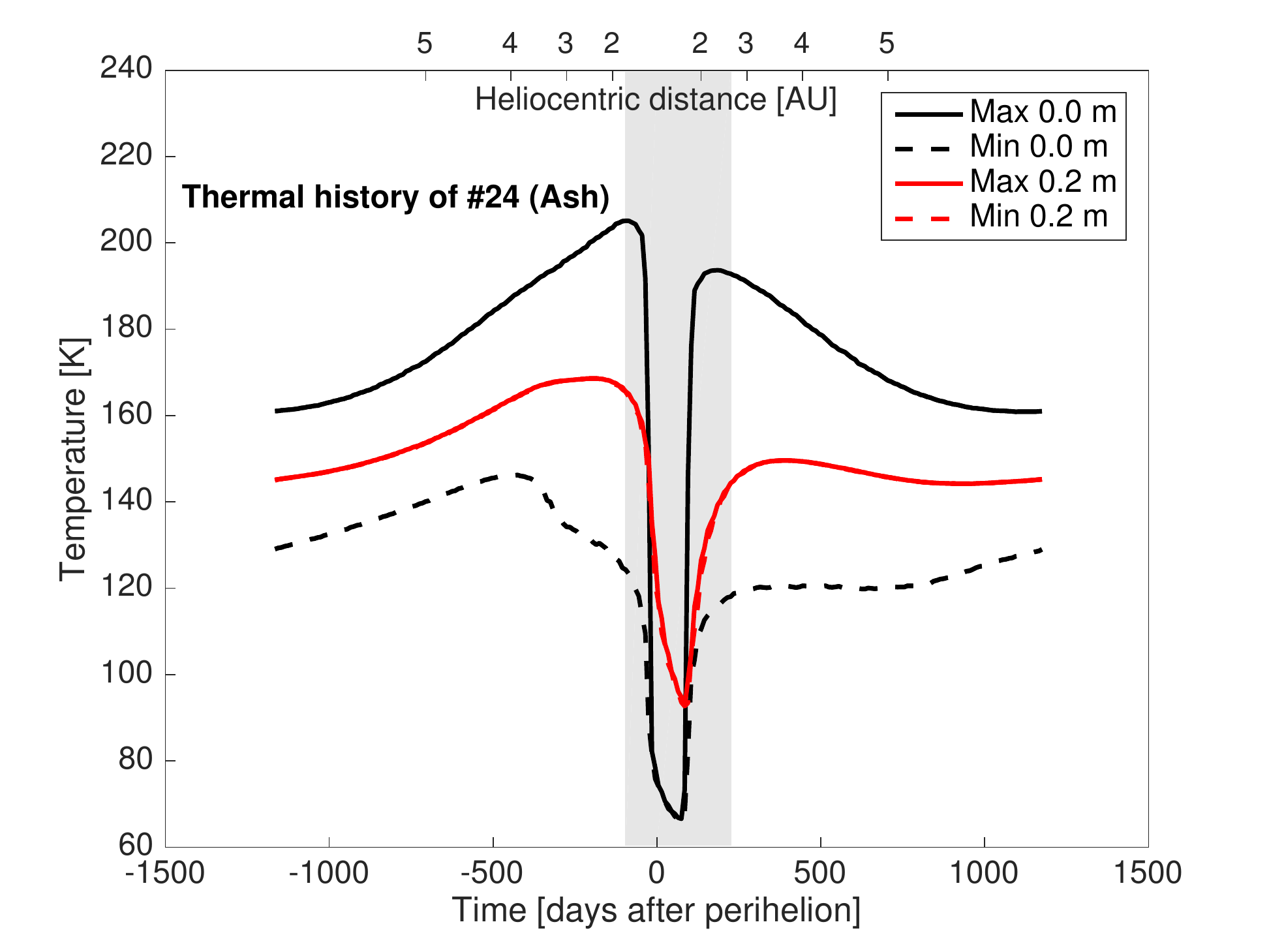}} & \scalebox{0.4}{\includegraphics{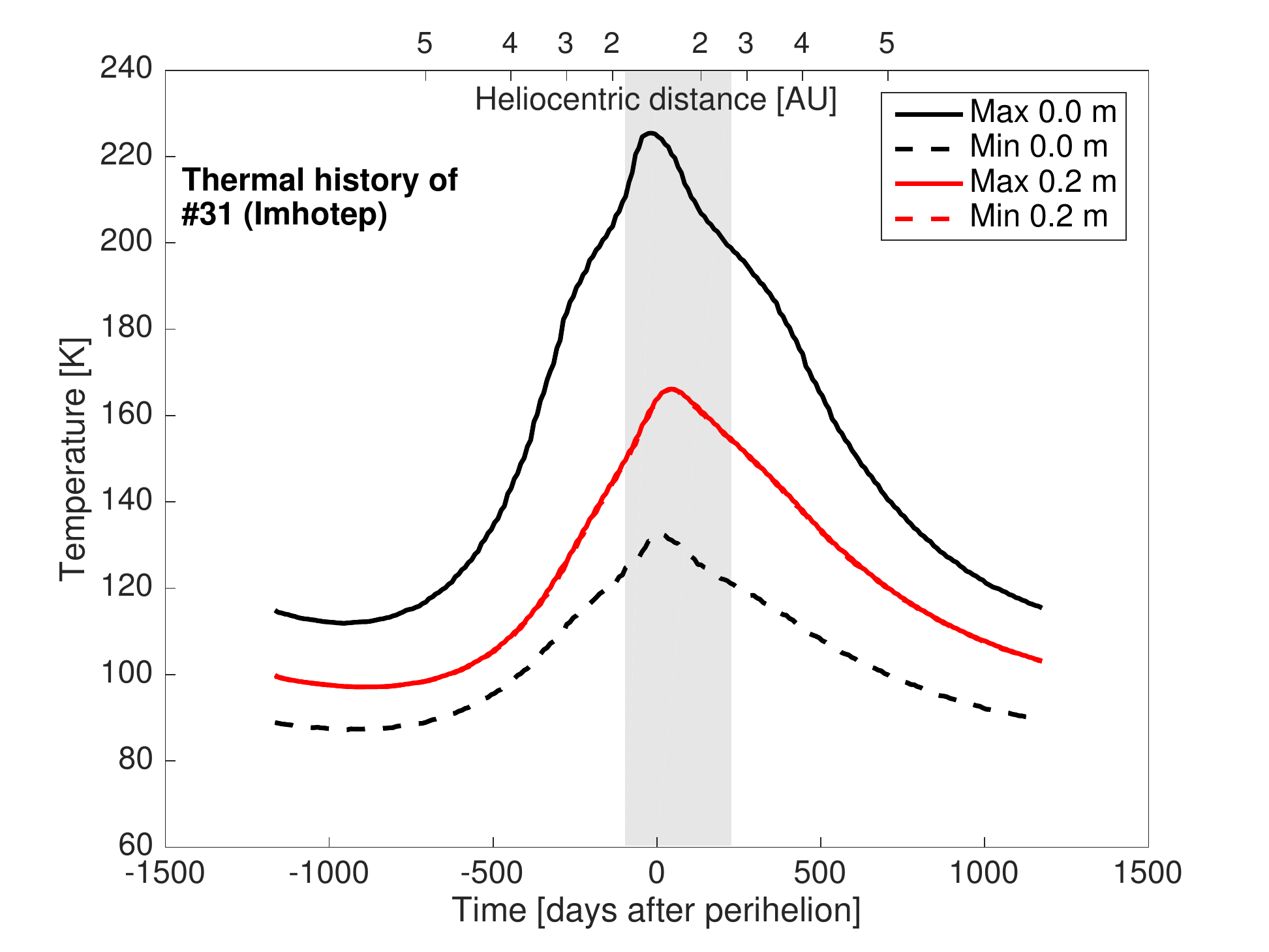}}\\
\end{tabular}
     \caption{Diurnal minimum and maximum temperatures as functions of time and heliocentric distance throughout the orbit of 67P at the surface and at $0.2\,\mathrm{m}$ below 
the surface for four target areas. The gray fields indicate the time period from the inbound equinox to the outbound equinox. Note that the red solid and 
dashed curves largely overlap. The panels to the right were first published, in a slightly different format, by \protect\shortciteN{takigawaetal19}.}
     \label{fig9}
\end{figure}

\noindent
The results of these simulations are shown in Fig.~\ref{fig9}. The black curves in those plots show the surface temperature, with the 
daytime peak as solid curves and the nighttime minimum as dashed curves. The gray areas mark the part of the orbit between the 
inbound and outbound equinoxes when the level of southern hemisphere activity is highest and most airfall takes place. Target areas 
\#3, \#16, and \#24 are all located on the northern hemisphere. Because of the spin axis orientation they are illuminated at aphelion and the 
surface temperatures increases as the comet travels toward perihelion. The temperature peaks around $200\,\mathrm{K}$ at all three 
locations, and the day/night amplitude is typically $10$--$30\,\mathrm{K}$. However, when the comet approaches the inbound equinox, 
the importance of the spin axis orientation becomes evident and the temperature plummets, as the regions enter into polar night. During the 
perihelion passage, the temperatures are as low as $60$--$70\,\mathrm{K}$ (with little to no dependence on nucleus rotational phase) and they 
are completely inactive. Therefore, airfall material originating from the strongly sublimating southern hemisphere has no problem landing in 
these target areas and they are flash--frozen after their coma flight. The generally good correspondence between inactivity and the deposition 
period in gray is the reason why we ignore self--cleaning for these areas.\\

\noindent
When target areas \#3, \#16, and \#24 emerge from polar night after the outbound equinox at $2.6\,\mathrm{AU}$, the surface temperatures are lower 
and (because of their strongly non--linear relation) the gas production rates are much lower than on the inbound branch, again due to the spin obliquity effect. 
Therefore, it is expected that much of the airfall ice collected at perihelion will survive the journey towards aphelion. Consistent with observations (e.g., \shortciteNP{gulkisetal15}; 
\shortciteNP{hassigetal15}; \shortciteNP{finketal16}), the northern hemisphere produces water vapor vigorously within $\sim 3\,\mathrm{AU}$ pre--perihelion because 
the airfall material in these regions reach $T\stackrel{>}{_{\sim}} 190$--$200\,\mathrm{K}$ for the first time since it was expelled from the southern hemisphere six years earlier.\\

\noindent
The situation for target area \#31, that is located on the southern hemisphere fairly close to the equator, is quite different.  It is 
poorly illuminated at aphelion, and its peak surface temperature is $\sim 40\,\mathrm{K}$ below the others. However, as the 
region approaches the Sun it remains fully illuminated and the temperature peaks at $\sim 230\,\mathrm{K}$, which is the highest 
among the four by far. Although it is bombarded with airfall material from even more active regions closer to the subsolar point at 
perihelion, it is capable of re--ejecting this material or preventing it from landing in the first place. That is why we considered self--cleaning 
for target areas \#30--31.\\

\noindent
Figure~\ref{fig9} also shows solid and dashed red curves. Those are the maximum and minimum temperatures $0.2\,\mathrm{m}$ below 
the surface. Because of the low conductivity of the porous comet material, the diurnal surface temperature amplitude is rapidly damped 
with depth, and at this level of magnification the solid and dashed curves are inseparable. In brief, $0.2\,\mathrm{m}$ below the surface, 
a thermometer would not be capable of telling whether the Sun is above or below the local horizon. We note that these calculations were 
performed without accounting for airfall. We ran one simulation for target area \#16 where a $0.5\,\mathrm{m}$ thick layer at temperature $200\,\mathrm{K}$ 
was deposited instantaneously onto the nucleus at perihelion when this location had a surface temperature of $\sim 70\,\mathrm{K}$. This procedure 
was intended to simulate the arrival of warm airfall (originating from the fully illuminated southern hemisphere) to the northern hemisphere that had 
polar night. After $\sim 150\,\mathrm{days}$ the temperature profile in the upper $0.2\,\mathrm{m}$ differed at most a few Kelvin between models 
with and without airfall. Because this thermal relaxation time is rather short compared to the orbital period we consider Fig.~\ref{fig9} a sufficiently 
good representation of temperatures at the target sites.\\

\noindent
A detailed comparison of outbound temperatures indicates that some regions will lose less of the airfall ice they have accumulated than 
others. In order to compare the sites with each other, we integrated the gas production rates from the outbound equinox to perihelion 
and normalized with respect to the site with the highest loss, thereby obtaining an index $L'$ quantifying the relative capacity to 
get rid of the ice. We then defined $S'=1-L'$ as the relative capacity to retain ice, and renormalized the set to the highest retainer. 
Thereby we obtained an index $\mathcal{S}$ that we refer to as ``survivability''. We also form an index $\mathcal{R}$ by normalizing 
the thickness of the accumulated airfall layer to the highest in the set that we call ``reachability''. If a target location simultaneously 
is characterized by a high reachability (it is reached by much airfall material) and high survivability (it remains comparably cool on the 
outbound orbital branch), then it is likely to be relatively active when the comet approaches perihelion on the inbound orbital branch.\\

\noindent
In this system, target site \#16 has $\{\mathcal{R},\,\mathcal{S}\}=\{1,\,1\}$ which means that eastern Ash has the potential 
of being more active before pre--perihelion than \#3 in Ma'at with  $\{\mathcal{R},\,\mathcal{S}\}=\{0.58,\,0.75\}$, and 
much more active than \#24 in western Ash ( $\{\mathcal{R},\,\mathcal{S}\}=\{0.11,\,0.31\}$) and \#31 in Imhotep 
( $\{\mathcal{R},\,\mathcal{S}\}=\{0.22,\,0.07\}$).

\section{Discussion} \label{sec_discussion}    

A number of authors have discussed the transfer of material from one location on 67P to another, the details of the airfall deposition 
process, and its effect on observable properties, using different methods. We here recapitulate their work and compare their results 
with our own. \shortciteN{pajolaetal17} calculated the maximum daily lift pressure (the product between production rate and expansion velocity that 
is proportional to the maximum size of liftable chunks) for different regions on 67P and used that information to offer an 
explanation for the differing size distributions at Agilkia (coarse and fine chunks) and Sais (coarse chunks with $D \stackrel{>}{_{\sim}} 3\,\mathrm{cm}$ only). Their discussion provides 
valuable insight into the dynamics of dust deposition and self--cleaning. The region Bes on the southern hemisphere reaches the highest 
lift pressure at perihelion and is expected to send a wide range of chunk sizes toward the northern hemisphere. Hapi experiences an 
extensive polar night period and can collect a substantial amount of material. Once illuminated, its own lift pressure is many orders of 
magnitude smaller than that of Bes, which, according to \shortciteN{pajolaetal17},  may explain why the coma in August 2014 and in the following months (primarily fed by Hapi contributions) 
was dominated by rather small ($0.1$--$1\,\mathrm{mm}$) chunks. Hapi is unable to eject most of the large chunks it receives and may 
experience an unusually large accumulation of airfall material over time, unless fragmentation of dm--m chunks into $0.1$--$1\,\mathrm{mm}$ chunks 
on the surface is efficient.  As a contrast, the lift pressure at Sais (once it emerges from polar night) is just an order of magnitude lower than for 
Bes, suggesting that all but the largest airfall chunks will be ejected after a brief residence on the nucleus. This may explain why the size distribution 
at that location (measured at the end of September 2016, $3.8\,\mathrm{AU}$ post--perihelion) was skewed towards coarse chunks. Agilkia, on the 
other hand, does not experience polar night at perihelion, which prevents it from accumulating chunks smaller than $\sim 0.1\,\mathrm{m}$ at 
that time. However, it was observed in November 2014 ($3.0\,\mathrm{AU}$ from the Sun, inbound), at a time when its lift pressure had 
been lower than that of Hapi since aphelion. The fine airfall seen at Agilkia but missing in Sais therefore originated from Hapi while its 
coarse airfall came from Bes and other regions on the southern hemisphere \shortcite{pajolaetal17}.\\

\noindent
In our work we only consider self--cleaning from regions \#30--31 that are strongly illuminated around perihelion. This means that 
we ignore the self--cleaning observed by \shortciteN{pajolaetal17} in, e.g., Sais at Ma'at, because the level of activity after 
the outbound equinox is still very low (Fig.~\ref{fig9}). Also, on the inbound leg we ignore the contribution arriving to our target 
areas from Hapi. In both cases, our assumptions concern very small particles, and we have already demonstrated that the 
mass carried by chunks smaller than a few centimeters is a relatively small fraction of the total mass (see Sec.~\ref{sec_method_airfall}). 
Therefore, our results are only slightly affected by this simplification (because we here only consider the total amount of mass, and 
do not attempt to accurately describe the size distribution of accumulated airfall at the target sites).\\

\noindent
\shortciteN{kramerandnoack15} performed numerical integration of dust chunk trajectories under the influence of realistic nucleus 
gravity, centrifugal and Coriolis forces (as they were working in a non--inertial frame), and gas drag. Their purpose was to explain the 
orientation of wind tails observed by \shortciteN{mottolaetal15} at Agilkia. They assumed that each point on the nucleus surface emitted 
the same fixed amount of gas flux. Thus, there was no dependence of local outgassing rate on solar location, the model nucleus had no 
day/night side, and the model coma was static and highly isotropic compared to the comet coma. However, the irregular nucleus shape created local 
concentrations of gas in strongly concave areas, like the Hapi neck region. The resulting dust flow had two major characteristics; 1) a flow out of Hapi 
that diverged into one flow over Seth and Ash on the large lobe and another flow over Hathor, Ma'at, and Hatmehit on the small lobe; and 2) a clockwise 
flow in the equatorial plane, as seen from the cometary north pole in a body--fixed frame, caused by the counter--clockwise nucleus rotation. 
Using this model, \shortciteN{kramerandnoack15} obtained dust flow vectors in the Agilkia region that were consistent with the wind tail directions 
observed by \shortciteN{mottolaetal15}.\\

\noindent
\shortciteN{laietal17} performed an investigation of airfall deposition that also considered realistic nucleus gravity, centrifugal and Coriolis forces, 
and gas drag. In their work, a fixed outgassing rate was applied to each point on the surface. These outgassing rates were taken to be 
proportional to the average cosine of the local day--time solar zenith angle during nucleus rotation. This anisotropic gas flow roughly captures 
the dependence of coma number density with nucleus latitude but smears its local time dependence, except for a strong day/night asymmetry. 
This anisotropic coma was merged at a 0.85 weight with an isotropic component with weight 0.15, and the total gas production rate was normalized to match the 
observed one. This was done once per month from January--December 2015 in order to account for the changing subsolar 
latitude and evolving total gas production rate with heliocentric distance. For each of the dozen static outgassing patterns, Direct Simulation Monte Carlo 
simulations were performed to calculate the gas number density, translational temperature, and expansion velocity as functions of position within the coma, needed 
for gas drag evaluations. \shortciteN{laietal17} used 14 dust classes with sizes in the $2.8\,\mathrm{\mu m}$--$4.5\,\mathrm{cm}$ range, each represented by 
$\sim 4\cdot 10^5$ particles launched from random locations on the surface that had different weights depending on local gas production conditions and chunk cross sections. 
In this manner, the local airfall and self--cleaning rates could be calculated at a given time, and the results were integrated over the considered orbital arc to 
estimate the local net deposition during one orbit. We note that \shortciteN{thomasetal15b} applied a similar model, however, limited the study to ejection of dust 
from Hapi to other parts of the northern hemisphere. They found that ejection speeds of $v_{\rm d}\stackrel{>}{_{\sim}} 0.5\,\mathrm{m\,s^{-1}}$ are necessary to 
leave the Hapi valley and that $\stackrel{>}{_{\sim}} 50\%$ of the particles escape the nucleus if $v_{\rm d}\stackrel{>}{_{\sim}} 1\,\mathrm{m\,s^{-1}}$, thus providing 
tight speed constraints for that transfer route.\\

\noindent
\shortciteN{laietal17} found that the largest liftable chunk on the southern hemisphere has $D=1\,\mathrm{cm}$ and that a net mass loss 
occurs everywhere except in Hapi where a net airfall deposit of $0.4\,\mathrm{m}$ depth is formed. We note that the $D\leq 1\,\mathrm{cm}$ 
limit is small compared to the observed numerous coma chunks in the $D=0.1$--$1\,\mathrm{m}$ class (e.g., \shortciteNP{rotundietal15}; \shortciteNP{davidssonetal15b}; 
\shortciteNP{agarwaletal16}), as well as compared to the size of chunks observed in resolved smooth terrains. Furthermore, their $0.5$--$1\,\mathrm{m}$ net loss on most of the northern 
hemisphere, combined with their $1$--$1.8\,\mathrm{m}$ net loss on most of the southern hemisphere, translates to a total loss of 
$\sim 2.5\,\cdot 10^{10}\,\mathrm{kg}$ (if applying an average loss of $1\,\mathrm{m}$, a nucleus surface area of $46.9\,\mathrm{km^2}$ and assuming that the 
density is that of the bulk nucleus, $530\,\mathrm{kg\,m^{-3}}$, see \shortciteNP{jordaetal16}). This is $2.4$ times higher than the observed nucleus net mass loss 
of $1.05(\pm 0.34)\cdot 10^{10}\,\mathrm{kg}$ according to the Rosetta/RSI 
radio science experiment \shortcite{patzoldetal19}. We suspect that the maximum liftable dust size at the perihelion subsolar point becomes significantly underestimated when calculating production rates 
based on the average cosine of the solar zenith angle. As a consequence, the amount of airfall is underestimated because the size distribution of ejected 
particles is truncated at a size below which return as airfall is highly uncommon. Simultaneously, the outgassing is rather strong in 
polar night regions (15\% of the daytime production), which leads to a high self--cleaning rate.\\

\noindent
Except for an initial acceleration phase aimed at determining a relation between local gas production rates, dust size, and their velocities, the current work does 
not consider gas drag effects on the dust orbits. This is admittedly a weakness. However, we do not think it is meaningful to engage in computationally 
demanding gas coma modeling that enables continuous gas drag, unless the nucleus model source that feeds the coma model is reproducing the 
strong temporal, latitudinal, and longitudinal variations in gas production rate known to occur on the real nucleus (during nucleus rotation and throughout the orbit). 
The summary above reminds us that such an elaborate treatment is beyond the current state--of--the--art of the field. It is unlikely that attempts to account for 
the orbital perturbations due to outgassing from the chunks would be realistic (due to their small size, chunks are likely having an excited spin state, and when 
combined with thermal inertia effects the time--evolution of the net non--gravitational force vector is unpredictable). Therefore, this ``rocket effect'' is ignored as well. 
Because of the point--mass assumption we also have not applied a realistic gravity field, but based on the comparison with a more appropriate approach described in 
Sec.~\ref{sec_method_numorb} we think that this may not have drastic consequences concerning our overall conclusions. Furthermore, we have ignored 
the effect of solar radiation pressure on the dynamics of coma chunks. Radiation pressure does have an affect on micron--sized chunks (e.g., \shortciteNP{tenishevetal11}), but considering the 
small contribution of such particles to airfall deposits in terms of mass, this simplification has no important effect on our conclusions.  With these limitations in mind, 
our calculations support the hypothesis (\shortciteNP{kelleretal15}; \citeyearNP{kelleretal17}) of a 
significant transport route from the southern to the northern hemisphere and demonstrate that both Ash and Ma'at are plausible destinations of airfall, 
as previously has been suggested (\shortciteNP{thomasetal15a}; \citeyearNP{thomasetal15b}). We find that the amount of airfall varies systematically between 
the four main regions of study (Ma'at, eastern and western Ash, and Imhotep) and within those regions. About 40\% of the considered target areas receive at 
most $0.5\,\mathrm{m}$ airfall and about $30\%$ receive at least $1\,\mathrm{m}$ according to our study. Considering the global coverage OSIRIS resolution of 
$0.2$--$3\,\mathrm{m\,px^{-1}}$ \shortcite{preuskeretal17} it is therefore not surprising that morphological changes because of airfall are difficult or 
impossible to verify observationally in many locations, but that such changes definitively are detected in others \shortcite{huetal17}.\\

\noindent
Our second goal consisted of estimating the amount of water ice loss during the transfer of chunks through the coma. We also wanted to explore the 
survivability of the substantially more volatile $\mathrm{CO_2}$ because of the chemical north/south dichotomy measured at 67P (\shortciteNP{hassigetal15}; 
\shortciteNP{fougereetal16}; \shortciteNP{finketal16}) and interpreted in the context of coma transport and airfall \shortcite{kelleretal17}. For a nominal set of model parameters 
our \texttt{NIMBUS} simulations showed that a dm--sized chunk would retain more than $90\%$ of its water ice when suspended for $12\,\mathrm{h}$ and that a 
cm--sized chunk would keep almost half its water ice in the same time. This lends substantial credibility to the claim by \shortciteN{kelleretal17} that the airfall is 
``wet'' and naturally explains the substantial level of water--driven activity on the northern hemisphere observed by Rosetta. We also found that a 
dm--sized chunk containing $5\%$ $\mathrm{CO_2}$ relative to water would lose all of it in $2\,\mathrm{h}$. It also means that a meter--sized chunk 
might be able to transfer supervolatiles from the south to the north. If so, then the low levels of $\mathrm{CO_2}$ outgassing from the northern 
hemisphere is a consequence of the small number of such bodies (and it may indicate that most $\mathrm{CO_2}$ ice could have been lost already prior to departure). 
Alternatively, some $\mathrm{CO_2}$ may have been transported from the south to the north, not as an independently condensed form of ice but trapped within water ice. 
Laboratory experiments (e.g., \shortciteNP{edridgeetal13}) demonstrate that a fraction of $\mathrm{CO_2}$ trapped in amorphous ice survives the release during 
crystallization as well as the cubic--hexagonal transformation and is released during water sublimation. We did not model that process explicitly. We note that Philae detected some 
highly volatile species at Agilika, such as $\mathrm{CO}$ and $\mathrm{CO_2}$ with Ptolemy \shortcite{wrightetal15}, and $\mathrm{CH_4}$ 
with COSAC \shortcite{goesmannetal15}. Those species may have been trapped in water ice within airfall material and released during water sublimation. Alternatively, they may have 
originated from underneath the airfall layer. \texttt{NIMBUS} simulations by \shortciteN{davidssonetal20} show that the measured pre--perihelion $\mathrm{CO_2}$ production rate curve is 
reproduced when the depth of the $\mathrm{CO_2}$ sublimation front is located $3.8\,\mathrm{m}$ below the surface on the northern hemisphere, and at $1.9\,\mathrm{m}$ below the 
surface on the southern hemisphere at aphelion (these depths are modified in the model over time because of dust mantle erosion as well as $\mathrm{CO_2}$ sublimation, see Sec.~\ref{sec_results_iceloss}). Supervolatiles 
like $\mathrm{CO}$ and $\mathrm{CH_4}$ may be released during sublimation of $\mathrm{CO_2}$ at the front, or from even larger depths by segregation processes, if they are 
trapped in $\mathrm{CO_2}$, as suggested by \shortciteN{gascetal17}. Although presence of $\mathrm{CO_2}$ in the airfall material itself remains a possibility, it cannot be the 
major source of $\mathrm{CO_2}$ release from the northern hemisphere before the inbound equinox, because the $\mathrm{CO_2}$ production does not correlate with that of $\mathrm{H_2O}$ 
(\shortciteNP{luspaykutietal15}; \shortciteNP{gascetal17}).\\

\noindent
Our third goal was to quantify the relative capacity of different airfall deposition sites to retain their water ice content after deposition. To this end we 
introduced a survivability index $\mathcal{S}$ and combined it with a reachability index $\mathcal{R}$ that measures the relative capability to collect airfall material. 
We performed an investigation limited to four sites, where \#16 in eastern Ash came out on top, with \#3 in Ma'at second, and \#24 (western Ash) and \#31 (Imhotep) on a shared third place. Comparing with the 
$\mathrm{H_2O}$ potential activity map of \shortciteN{fougereetal16} derived from long measurement series by ROSINA and corrected for illumination biases, it is
 clear that \#16 is substantially more active than \#24 and \#31, which is consistent with our work. However, \#3 is the most active of them all. Interestingly, 
the high--activity belt shown by \shortciteN{fougereetal16}, stretching from Ash, Seth, to Ma'at along the (anti)meridian with an epicenter in Hapi, has a strong 
resemblance with the simulated model flow pattern of airfall material originating from Hapi (\shortciteNP{thomasetal15b}; \shortciteNP{kramerandnoack15}). 
It is therefore possible that the ice variability introduced by airfall from the southern hemisphere at perihelion is mixed and masked by redistribution of airfall due to 
Hapi activity at the time ROSINA performed the measurements. It is encouraging that the activity map of \shortciteN{laraetal15}, based on the footprint of dust 
jets observed early during the Rosetta mission, suggest high levels of activity both at \#3 and \#16.\\

\begin{figure}
\begin{center}
     \scalebox{0.55}{\includegraphics{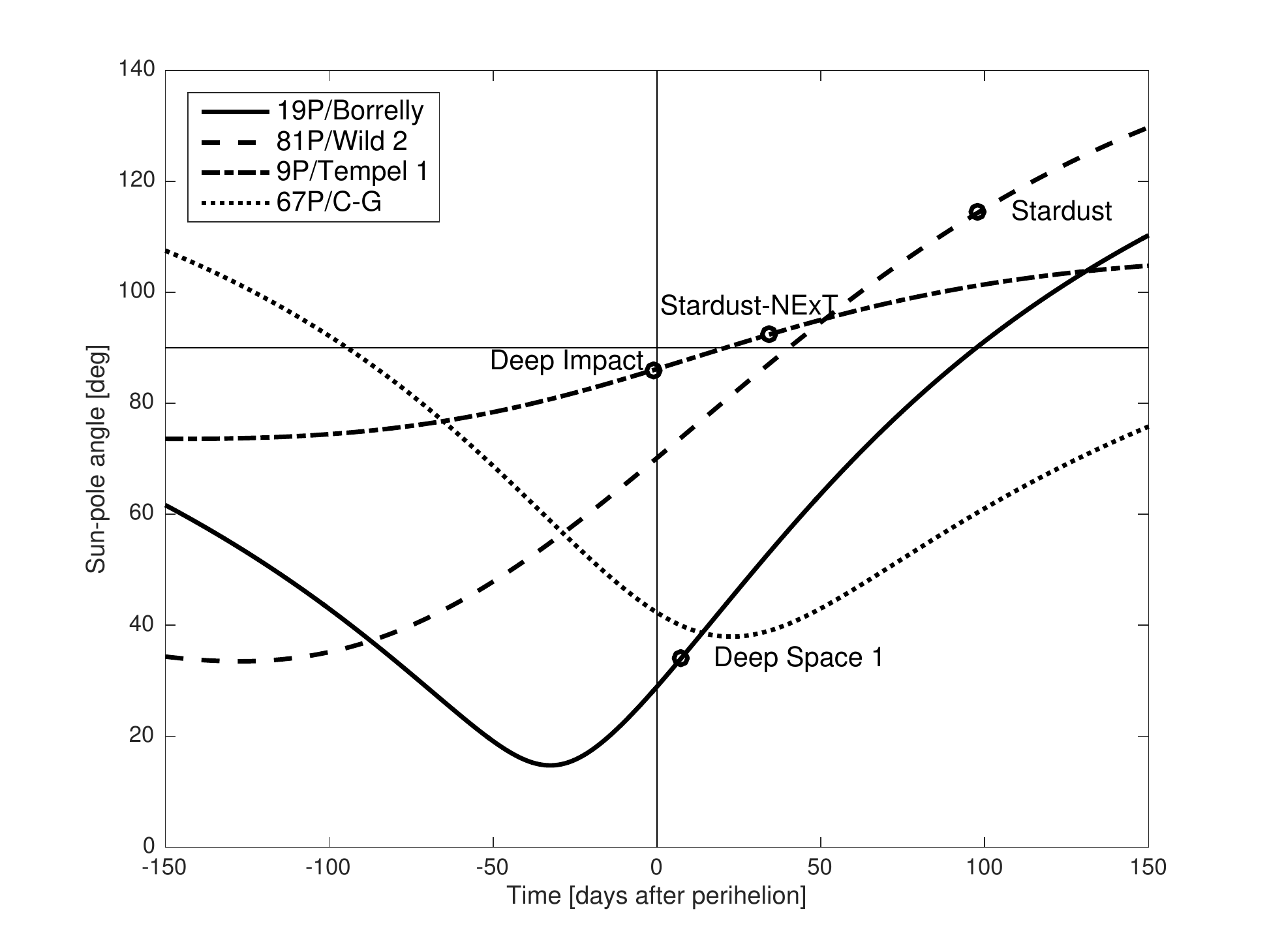}}
   \caption{Angle between the pole facing sunward at perihelion and the comet--Sun vector versus time. Spacecraft flyby dates are indicated. Perihelion 
is marked with a vertical line. The airfall deposit process is least efficient on the horizontal line (no polar night).}
     \label{fig11}
\end{center}
\end{figure}

\noindent
To what extent are airfall deposits common features on comet nuclei? We expect that three criteria need to be fulfilled in order to create thick and wide--spread 
airfall deposits. First, the spin pole needs to have such an orientation that a large fraction of the nucleus surface experiences polar night at perihelion. In terms of 
the spin axis obliquity and argument (for definitions see, e.g., \shortciteNP{sekanina81}; \shortciteNP{davidssongutierrez04}) the most favorable configuration 
is an obliquity near $90^{\circ}$ and an argument near $90^{\circ}$ or $270^{\circ}$ (i.e., the perihelion subsolar point coincides with one of the comet nucleus poles). 
Second, the post--perihelion equinox should take place as late as possible. The equinox passage means that the Sun is in the equatorial plane of the comet and that 
fresh airfall deposits no longer are protected from solar illumination. If airfall is illuminated sufficiently close to perihelion, strong activity and self--cleaning may take 
place that prevents long--term deposits from forming.  Third, the comet needs to produce abundant chunks that are too large to reach escape velocity. The presence of 
a detectable debris trail is a potential indicator that a specific comet is prone to eject particularly large chunks. A Spitzer survey of 34 comets showed that 27 objects, or $79\%$, 
had trails consisting of millimeter--sized particles and larger \shortcite{reachetal07}, suggesting that 67P is not unusual.\\

\noindent 
Figure~\ref{fig11} shows the time evolution of the angle between the solar direction and the comet nucleus pole that is sunlit at perihelion (here referred to as the polar angle), 
for four comets that were targets of spacecraft missions. The smaller the polar angle, the larger the fraction of the comet surface has polar night at perihelion. 
If it reaches $90^{\circ}$ (marked by the horizontal line), equinox takes place and the Sun is in the equatorial plane of the comet (i.e., no region has polar night). Among the four, Comets 67P and 19P/Borrelly 
seem to have the best conditions for airfall deposition. The perihelion polar angles are $42^{\circ}$ for 67P and  $29^{\circ}$ for 19P/Borrelly, i.e., the polar night regions are large. 
Comet 67P reaches equinox later (221 days post--perihelion, outside the figure) than 19P/Borrelly (97 days post--perihelion), but self--cleaning is probably limited in both cases. 
67P evidently produces large chunks (as observed by Rosetta, but also inferred from the existence of a trail; \shortciteNP{sykeswalker92}; \shortciteNP{kelleyetal09}; \shortciteNP{ishiguroetal09}), 
that enables airfall deposit formation, but this may not be the case for 19P/Borrelly. The reports on the Deep~Space~1 flyby of 19P/Borrelly do not mention large coma chunks 
(\shortciteNP{soderblometal02}; \shortciteNP{boiceetal02b}), and the comet does not have a trail \shortcite{ishiguroetal09}. Despite the favorable geometric conditions, Comet 19P/Borrelly 
may therefore not have large smooth terrains. Unfortunately, Deep Space 1 images cannot provide an answer because the flyby took place just 8 days after perihelion \shortcite{soderblometal02}, 
i.e., a potential airfall coverage would have been hidden from view on the dark polar night side.\\

\noindent
Comets 81P/Wild~2, and particularly 9P/Tempel~1, are least suitable for vast airfall deposit formation. 9P/Tempel~1 has a perihelion polar angle of $86^{\circ}$ and passes 
equinox just three weeks later. It is therefore surprising that 9P/Tempel~1 has prominent smooth areas \shortcite{ahearnetal05b}. Particularly the smooth area ``S2'' near the south pole 
\shortcite{veverkaetal13} would be most consistent with airfall onto the weakly illuminated and least active part of the nucleus, while the origin of localized smooth areas elsewhere is less obvious. 
It is also interesting to note that observations of a slowly expanding coma arc during the Deep~Impact flyby suggests the presence of fairly large (millimeter--sized) slow--moving chunks \shortcite{farnhametal07}, 
and a trail associated with Comet 9P/Tempel~1 has been observed both with IRAS  \shortcite{sykeswalker92} and with Spitzer \shortcite{reachetal07}. This suggests that the spin axis orientation 
plays a minor role for the existence of smooth airfall terrain, though a large coverage probably still requires a small perihelion polar angle. If true, this could mean that all comets with an appropriate 
dust size distribution are capable of building smooth airfall terrains. For that reason, the case of 81P/Wild~2 is somewhat puzzling. 
With a polar angle of $70^{\circ}$, a rather large part of the nucleus has polar night at perihelion. The equinox passage is just 40 days post--perihelion, but judging from Comet 9P/Tempel~1, 
that may not prevent airfall debris from remaining on the surface. The Stardust spacecraft collided with a couple of mm--sized dust grains \shortcite{tuzzolinoetal04} and the comet 
has a trail according to \shortciteN{ishiguroetal09}. Furthermore, the Stardust flyby took place 98 days post--perihelion \shortcite{brownleeetal04}, at a time when the regions that 
experienced polar night at perihelion had become visible. Yet, the Stardust images \shortcite{brownleeetal04} do not reveal smooth terrains of the type seen on 9P/Tempel~1 or like Imhotep or Hapi on 67P. 
Potentially, the airfall material on 81P/Wild~2 is hiding in plain view, i.e., as a thin coverage on top of the underlying rugged topography, similar to the Ash and Seth regions on 67P. 
The nature of those regions only became evident in Rosetta images that had significantly better resolution than the best Stardust images ($14\,\mathrm{m\,px^{-1}}$;  \shortciteNP{brownleeetal04}). 
Alternatively, airfall deposits do not form on Comet 81P/Wild~2.\\

\noindent
Two spacecraft targets, Comets 1P/Halley and 103P/Hartley~2, were omitted from Fig.~\ref{fig11} because they are complex rotators (the polar angle is not easily calculated). Large chunks 
are common around 103P/Hartley~2 \shortcite{kelleyetal13} and it has a trail \shortcite{reachetal07}. EPOXI images revealed a smooth neck region \shortcite{ahearnetal11} reminiscent  
of Hapi on 67P. With airfall confirmed on 67P, and suspected on 9P/Tempel~1 and 103P/Hartley~2, it indeed appear to be a common process in comets.

\section{Conclusions} \label{sec_conclusions}

We have modeled the transfer of material from the highly active southern hemisphere of comet 67P to its northern hemisphere during 
the perihelion passage. We find that such transport routes exist and that the Ash and Ma'at regions receive plenty of airfall, thereby supporting 
previous hypotheses (\shortciteNP{thomasetal15a}, \citeyearNP{thomasetal15b}; \shortciteNP{kelleretal15}, \citeyearNP{kelleretal17}). 
The amount of airfall accumulated between the inbound and outbound equinoxes is typically a few times $0.1$--$1\,\mathrm{m}$ (some of which 
will be removed in other parts of the orbit, i.e., these numbers should not be interpreted as \emph{net} orbital accumulations, but rather 
as an estimate of the material the northern hemisphere has to its disposal to drive activity pre--perihelion). The distribution of 
airfall is heterogeneous, with the most substantial accumulation in eastern Ash, followed by western Ash and Ma'at at similar levels, with 
the least airfall at central Imhotep because of efficient self--cleaning (only these four regions were considered, e.g., Hapi was not included in the study).\\

\noindent
We have also modeled the loss of both $\mathrm{H_2O}$ and $\mathrm{CO_2}$ from $0.01$--$0.1\,\mathrm{m}$ diameter chunks in 
the coma, using the elaborate \texttt{NIMBUS} model. We find that a cm--sized chunk can retain roughly half its water ice during 
a $12\,\mathrm{h}$ exposure to the Sun at perihelion, while a dm--sized chunk holds on to more than 90\% of the water ice. If there is 
$\mathrm{CO_2}$ at a 5\% level relative to water by number, a dm--sized chunk loses all carbon dioxide in two hours. We therefore 
support the scenario described by \shortciteN{kelleretal17} of a ``wet'' airfall deprived of supervolatiles that is responsible for the 
observed water--dominated comet activity prior to the inbound equinox.\\

\noindent
Finally, we studied the longterm loss of ice following the near--perihelion airfall accumulation. We demonstrated that the surface 
temperatures are kept comparably cool on the northern hemisphere during the outbound orbital branch, which helps explain 
the high level of activity in the north on the inbound orbital branch when the deposits finally are heated to high temperature. 
We introduce the reachability and survivability indices $\mathcal{R}$ and $\mathrm{S}$ as a way to classify the potential of 
a region to have a high level of activity by simultaneously accumulating a high amount of wet airfall and to preserve it to the 
following perihelion passage. This work also serves as a guide for selecting the most promising volatile 
rich smooth terrain sites in the northern hemisphere for future comet nucleus sample return missions.

\bigskip
\bigskip

\noindent
{\sl ACKNOWLEDGMENTS.}  Parts of the research was carried out at the Jet Propulsion Laboratory, California Institute of Technology, under a contract with the National Aeronautics and Space Administration.\\

\noindent
\emph{COPYRIGHT}.  \textcopyright\,2020. All rights reserved.

\bibliography{Davidsson_etal_airfall_ver04acc.bbl}

\clearpage \linespread{1}

\end{document}